\documentclass[a4paper,11pt]{article}
\pdfoutput=1 
\usepackage{jcappub} 
\usepackage{amsmath,amsfonts,amssymb,mathtools}
\usepackage[numbers]{natbib}
\usepackage[T1]{fontenc} 

\renewcommand{\vec}{\mathbf}

\newcommand{\be}{\begin{equation}}
\newcommand{\ee}{\end{equation}}
\newcommand{\bea}{\begin{eqnarray}}
\newcommand{\eea}{\end{eqnarray}}
\newcommand{\bdm}{\begin{displaymath}}
\newcommand{\edm}{\end{displaymath}}
\newcommand{\nn}{\nonumber}

\newcommand{\xv}{\mathbf{x}}

\newcommand{\kv}{\mathbf{k}}

\newcommand{\qv}{\mathbf{q}}
\newcommand{\pv}{\mathbf{p}}

\allowdisplaybreaks

\title{\boldmath Wide-angle effects in the galaxy bispectrum}
\author[a,b,c,d,e,1]{Kevin Pardede,\note{Corresponding author.}}
\author[f]{Enea Di Dio}
\author[g]{ and Emanuele Castorina}

\affiliation[a]{SISSA - International School for Advanced Studies, Via Bonomea 265, 34136 Trieste,  Italy}
\affiliation[b]{ICTP, International Centre for Theoretical Physics, Strada Costiera 11, 34151, Trieste, Italy}
\affiliation[c]{Institute for Fundamental Physics of the Universe, Via Beirut 2, 34151 Trieste, Italy}
\affiliation[d]{Istituto Nazionale di Fisica Nucleare, Sezione di Trieste,  via  Valerio  2,  34127 Trieste,  Italy}
\affiliation[e]{Istituto Nazionale di Astrofisica, Osservatorio Astronomico di Trieste, via Tiepolo 11, 34143 Trieste, Italy}
\affiliation[f]{Theoretical Physics Department, CERN, 1211 Geneva 23, Switzerland}
\affiliation[g]{Dipartimento di Fisica ‘Aldo Pontremoli’, Universita’ degli Studi di Milano, \\ Via Celoria 16, 20133 Milan, Italy}

\emailAdd{kpardede@sissa.it}

\abstract{Primordial non-Gaussianities (PNG) leave unique signatures in the bispectrum of the large-scale structure. With upcoming galaxy surveys set to improve PNG constraints by at least one order of magnitude, it is important to account for any potential contamination. In our work we show how to include wide-angle effects into the 3-dimensional observed galaxy bispectrum. We compute the leading wide-angle corrections to the monopole, finding that they could mimic local PNG with an amplitude of $f_{\rm NL} = \mathcal{O}\left( 0.1 \right)$. We also compute the dipole induced by wide-angle effects, whose amplitude is a few-percent of the flat-sky monopole.
We estimate that wide-angle effects in the monopole can be safely neglected for survey volumes of the order of $8~\mathrm{Gpc}^3 h^{-3}$, while the dipole can start being detected from surveys probing volumes larger than $50~\mathrm{Gpc}^3 h^{-3}$. Our formalism can be readily adapted to realistic survey geometries and to include relativistic effects, which may become relevant at high redshifts.}

\begin{document}

  \begin{minipage}{.45\linewidth}
    \begin{flushleft}
    \end{flushleft}
  \end{minipage}
\begin{minipage}{.45\linewidth}
\begin{flushright}
 {CERN-TH-2023-033}
 \end{flushright}
 \end{minipage}

\maketitle
\flushbottom

\section{Introduction}
\label{sec:intro}

The large-scale structure (LSS) of the Universe encodes a wealth of cosmological information, which is typically compressed in a small number of summary statistics.
For a Gaussian field, all the information is encoded in the power spectrum, i.e.~the Fourier transform of the 2-point correlation function.
Non-Gaussianities, either produced by the non-linear gravitational interaction or set in the initial conditions provided by inflation, generates non-vanishing 3-point functions. Therefore the study of its Fourier transform, the bispectrum, provides complementary information to the power spectrum, in particular on the non-linear evolution~\cite{Fry1994B, HivonEtal1995, MatarreseVerdeHeavens1997, ScoccimarroEtal1998, Scoccimarro2000B, SefusattiEtal2006, GagraniSamushia2017, YankelevichPorciani2019, OddoEtal2019, DAmicoEtal2020, GualdiVerde2020, HahnEtal2020, AlkhanishviliEtal2021, OddoEtal2021,  GualdiEtal2021, HahnVillaescusaNavarro2021, AgarwalEtal2021,  IvanovEtal2021A, DAmicoEtal2022C, RizzoEtal2022, BiagettiEtal2022} and on the dynamics of Inflation~\cite{ScoccimarroEtal2001B, ScoccimarroSefusattiZaldarriaga2004, SefusattiKomatsu2007, JeongKomatsu2009B, Sefusatti2009, 
TasinatoEtal2014, TellariniEtal2015, TellariniEtal2016, MoradinezhadDizgahEtal2018, KaragiannisEtal2018, CastorinaMoradinezhad2020, MoradinezhadDizgahEtal2021, Barreira2022}. 

Analyzing measurements of the bispectrum, however, presents a number of technical challenges, and the first analyses of the BOSS data~\cite{BOSS:2016wmc} have been attempted only recently~\cite{Gil-Marin2014A, Gil-Marin2014B, Gil-Marin2016, DAmicoEtal2020,PhilcoxIvanov2021, CabassEtal2022A, DAmicoEtal2022, CabassEtal2022B, PhilcoxEtal2022, DAmicoEtal2022B, IvanovEtal2023}.
An example of the intrinsic complication in a bispectrum analysis is provided by finite volume effects convolution, recently studied in \cite{SugiyamaEtal2018, Philcox2020,Philcox2021,PardedeEtAl2022,AlkhanishviliEtal2022, IvanovEtal2023}. Additionally, all the analyses so far have been working in the so-called flat-sky, or distant observer, approximation, by considering the lines of sight to different galaxies to be parallel. 

In this work we show how to go beyond the flat-sky approximation in the bispectrum analysis, by expanding perturbatively in the ratio between the pair separations and the surveys effective comoving distance. Departures from the flat-sky regime are usually referred to as wide-angle effects.
Given the large fraction of the sky covered by current and upcoming galaxy surveys such as DESI \cite{DESI:2013agm}, \textit{Euclid} \cite{EUCLID:2011zbd} and SPHEREx \cite{Dore:2014cca}, wide-angle effects need to be quantified to make sure that the analyses are robust.

Recently, similar questions were addressed by the Authors of \cite{NoorikuhaniScoccimarro2022}, who computed wide-angle corrections the the bispectrum multipoles using a Cartesian basis. In this work we will provide a different formulation of wide-angle effects, based on an expansion in spherical tensors, which is more easily adapted to the state-of-the art approaches to the convolution of the theoretical model with a survey window function \cite{PardedeEtAl2022}, whose effects become important on the same scales of the wide-angle corrections.

For the power spectrum, wide-angle effects have been studied extensively \cite{Hamilton1992ApJ, HamiltonCulhane1996, ZaroubiHoffman1996, Szalayetal1998, Bharadwaj1998, Szapudi2004, PapaiSzapudi2008, ShawLewis2008, Raccanelli:2013dza, YooSeljak2013, SlepianEisenstein2015, Reimbergeatal2015,CastorinaWhite2017,CastorinaWhite2018,CastorinaWhite2019,PhilcoxSlepian2021,ElkhashabEtAl2021,NoorikuhaniScoccimarro2022}. As the consequence of the partially broken translation invariance, the power spectrum is no longer diagonal, $\langle\delta(\kv_1)\delta(\kv_2) \rangle = P(\kv_1, \kv_2)$. The current  approach in the community is to consider the so called local estimator (e.g.~$P(\kv, \xv)$ for the power spectrum, where $\hat{\xv}$ is the line of sight vector), where "local" here refers to a small region where the translation invariance is approximately preserved \cite{Scoccimarro2015}. Wide-angle effects then can be captured perturbatively by expanding over the small parameter\footnote{The expansion parameter $(k x)^{-1}$ is the Fourier counterpart of the ratio between the pair separation over the survey effective comoving distance.} $\left(k x \right)^{-1}$ (see e.g. \cite{CastorinaWhite2017, BeutlerCastorinaZhang2018})
\begin{align}
P(\kv, \xv) &= \sum_{n} P^{(n)}(\kv, \hat{\xv}) \left( k x \right)^{-n}.
\end{align}

In principle the same technique can be applied to the bispectrum. As recently shown by \cite{NoorikuhaniScoccimarro2022}, under the assumption of periodic boundary condition (neglecting window function effects), the wide-angle effects in the bispectrum can be described perturbatively, written schematically as follows 
\begin{align}
B(\kv_1, \kv_2, \xv) &= { \sum_n}\sum_{i+j = n} B^{(i+j)}(\kv_1, \kv_2, \hat{\xv}) \left(k_1 x \right)^{-i} \left(k_2 x \right)^{-j}.
\end{align}
In this work we are interested in providing an alternative formulation of wide angle effects in the bispectrum multipoles for the commonly used Scoccimarro estimator \cite{Scoccimarro2015}, which we think is more suitable for the convolution with the survey window function.

Measurements of the large scale bispectrum offer a unique window into the Early Universe as revealed by the possible presence of  Primordial Non-Gaussianity (PNG) \cite{Achucarro:2022qrl}. The wide-angle corrections to the monopole of the power spectrum and of the bispectrum have a similar parametric expression to the amplitude of local PNG, and could therefore be misinterpreted for a detection of PNG in LSS data. 
We will show that wide-angle effects produce an equivalent amplitude of local PNG with $f_{\rm NL} \sim 0.2$.

Beyond the flat-sky approximation, odd multipoles are generated by wide-angle corrections. In the recent literature~\cite{Clarkson:2018dwn,Jeong:2019igb,deWeerd:2019cae}, the dipole of the bispectrum has received some attention, since it can be sourced by relativistic effects as well. Similarly to what happens for the dipole of the power spectrum ~\cite{Bonvin:2018ckp, Bonvin:2020cxp, Castello:2022uuu, Sobral-Blanco:2022oel, Jimenez:2022evh}, the latter offer the opportunity to test modifications to the Euler equation with a measurement of the odd multipoles of the bispectrum. Moreover, the bispectrum is the lowest-order statistic, for a single tracer, sensitive to parity violating physics. 
Therefore, the understanding of any other source of odd multipoles is crucial in the search for parity-odd new physics in galaxy clustering, such as vector-type parity-violation~\cite{Philcox:2022hkh, HouSlepian2022}. Here we derive the dipole induced by wide-angle corrections,  which we expect to be of a similar magnitude, or greater, than the one induced by relativistic effects.
Our formalism can be easily extended to incorporate relativistic corrections as well.

This work is organized as follows. The wide-angle effects formulation for the bispectrum multipoles is laid out in Section \ref{sec:wa_in_bisp}. We present the numerical implementation of our approach in Section \ref{sec:num_results}. Finally, we give our conclusions and outlook in Section \ref{sec:conclusion}.

\section{Wide angle in the Bispectrum multipoles estimator}
\label{sec:wa_in_bisp}
In this section we introduce the bispectrum estimator and we compute its ensemble average by accounting for wide-angle effects.
We start by considering the Scoccimarro bispectrum estimator~\cite{Scoccimarro2015} 
\begin{align}
\hat{B}_\ell(k_1,k_2,k_3) &= \frac{2\ell+1}{V_B V} \prod_{i=1}^3 \left[\int_{k_i} d^3 q_i  \int_V d^3 x_i e^{-i \qv_i \cdot \xv_i}\right] ~ \delta_D(\qv_{123})  \delta_g(\xv_1) \delta_g(\xv_2) \delta_g(\xv_3)  \mathcal{L}_\ell(\hat{\qv}_1 \cdot \hat{\xv}_3),
\end{align}
where the integration $\int_k d^3 q$ runs over the spherical shell of radius $q$ in the range $k-\Delta k/2 \leq q \leq k + \Delta k/2$. The fluctuation in the galaxy number density is defined as $\delta_g\left( \xv \right)$ and we have introduced
\begin{equation}
    V_B \equiv V_B(k_1, k_2, k_3) \equiv \int_{k_1} d^3 q_1 \int_{k_2} d^3 q_2 \int_{k_3} d^3 q_3~\delta_D(\qv_{123}).
\end{equation}
We are interested in computing the ensemble average of the bispectrum estimator without relying on the flat-sky approximation.
Hence, ignoring the $k$-binning, we have
\begin{align}
\label{eq:ensamble_bisp}
\langle \hat{B}_\ell(k_1,k_2,k_3) \rangle&= (2 \ell+1) \int  \frac{d^3 x_3}{V} \int d^3 x_{13} \int d^3 x_{23} ~ \langle \delta_g(\xv_1) \delta_g(\xv_2) \delta_g(\xv_3) \rangle 
\nn \\
& \hspace{5cm}
e^{-i \kv_1 \cdot \xv_{13}} e^{-i \kv_2 \cdot \xv_{23}} \mathcal{L}_\ell(\hat{\kv}_1 \cdot \hat{\xv}_3)\, ,
\end{align}
where we have introduced the variables
\be
\xv_{13}= \xv_1 - \xv_3\, ,\quad
\xv_{23}= \xv_2 - \xv_3 \, .
\ee
From Eq.~\eqref{eq:ensamble_bisp} it is clear that we need to compute the 3-point correlation function (3PCF) beyond flat-sky.
Therefore, we consider the (deterministic part of) galaxy number density $\delta_g(\vec{k})$, perturbatively expanded up to second order in the matter density field $\delta \left( \mathbf{k} \right)$
\begin{align}
    \delta_g(\kv,\hat{\xv}) = Z_1(\kv, \hat{\xv}) \delta^{(1)}(\kv) + \int \frac{d^3 q}{(2\pi)^3} Z_2(\qv, \kv - \qv, \hat{\xv})\delta^{(1)}(\qv)\delta^{(1)}(\kv-\qv),
\end{align}
where we have made the dependence on the line of sight $\hat{\xv}$ explicit in the kernels~\cite{Bernardeau:2001qr}
\begin{align} 
\label{eq:Z1}
    Z_1(\kv, \hat{\xv}) &= b_1 + f (\hat{\kv} \cdot \hat{\xv})^2, 
    \\
    \label{eq:Z2}
    Z_2(\kv, \hat{\xv}) &= b_1 F_2(\kv_1, \kv_2) + \frac{b_2}{2} + b_{\mathcal{G}_2} S(\kv_1, \kv_2) + f \mu_{12}^2 G_2(\kv_1, \kv_2) 
    \\ \nonumber & \hspace{2cm}+ \frac{f k_{12} \mu_{12}}{2} \left[\frac{\hat{\kv}_1 \cdot \hat{\xv}}{k_1} Z_1(\kv_2, \hat{\xv})  + \frac{\hat{\kv}_2 \cdot \hat{\xv}}{k_2} Z_1(\kv_1, \hat{\xv}) \right],  
\end{align}
with $b_1, b_2, b_{\mathcal{G}_2}$ are the linear, quadratic, bias related to second order Galileon field $\mathcal{G}_2$ respectively, $\kv_{12} \equiv \kv_1 + \kv_2$ and $\mu_{12} \equiv \hat{\kv}_{12} \cdot \hat{\xv}$. The functions $F_2$ and $G_2$ are the standard second order matter density and velocity quadratic kernels~\cite{Bernardeau:2001qr}, and
\begin{equation}
    S(\kv_1, \kv_2) = (\hat{\kv}_1 \cdot \hat{\kv}_2)^2 - 1\,.
\end{equation}
In full generality, the tree-level 3-point function can be written as
\begin{align}
\label{3pcf_beyond_pp}
    \langle \delta(\xv_1) \delta(\xv_2) \delta(\xv_3) \rangle &= \int \frac{d^3 p_1}{(2 \pi)^3} \frac{d^3 p_2}{(2 \pi)^3} \frac{d^3 p_3}{(2 \pi)^3} (2\pi)^3 \delta_D(\pv_{1} + \pv_{2} + \pv_{3}) e^{i \pv_1 \cdot \xv_1} e^{i \pv_2 \cdot \xv_2} e^{i \pv_3 \cdot \xv_3}
    \nn \\& \hspace{1cm} 2\Big[Z_1(\pv_1, \hat{\xv}_1) Z_1(\pv_2, \hat{\xv}_2) Z_2(\pv_1, \pv_2, \hat{\xv}_3) P(p_1) P(p_2) 
    \nn \\& \hspace{3cm} + Z_1(\pv_2, \hat{\xv}_2) Z_1(\pv_3, \hat{\xv}_3) Z_2(\pv_2, \pv_3, \hat{\xv}_1) P(p_2) P(p_3) 
    \nn \\& \hspace{3cm} + Z_1(\pv_3, \hat{\xv}_3) Z_1(\pv_1, \hat{\xv}_1) Z_2(\pv_3, \pv_1, \hat{\xv}_2) P(p_3) P(p_1) 
    \Big] 
    \nn \\&= \int \frac{d^3 p_1}{(2\pi)^3} \frac{d^3 p_2}{(2\pi)^3} e^{i \pv_1 \cdot \xv_{13}} e^{i \pv_2 \cdot \xv_{23}} \mathcal{F}(\pv_1, \pv_2, \hat{\xv}_1, \hat{\xv}_2, \hat{\xv}_3),
\end{align}
where we have collected the kernels into
\begin{align}
\label{F_kernels}
    \mathcal{F}(\pv_1, \pv_2, \hat{\xv}_1, \hat{\xv}_2, \hat{\xv}_3) = 2 Z_1(\pv_1, \hat{\xv}_1) Z_1(\pv_2, \hat{\xv}_2) Z_2(\pv_1, \pv_2, \hat{\xv}_3)  P(q_1) P(q_2)+ \mathrm{perms.}
\end{align}
Since all the angular dependencies are analytical\footnote{Due to the mode-coupling between $\pv_1$ and $\pv_2$, the integrals over $\mathcal{L}_{\ell_3} \left(\hat{\pv}_1 \cdot \hat{\pv}_2\right) $ needs to be computed numerically. We discuss the convergence of the sum over $\ell_3$ in the Appendix~\ref{sec:convergence_test}.}, we can evaluate the 3PCF by expanding the 
$\mathcal{F}$ kernel in terms of two wide-angle expansion parameters\footnote{As shown in Ref.~\cite{Castorina:2021xzs}, for the monopole this perturbative expansion at second order in terms of the ratio of the pair separation over the line-of-sight comoving distance is accurate at the percent level up to the boundary of the survey window function.} $x_{13}/x_3$ and $x_{23}/x_3$ (where $\hat {\xv}_3$ has been chosen to be the line-of-sight)
\begin{align}
\label{F_in_wa_params}
    \mathcal{F}(\pv_1, \pv_2, \hat{\xv}_1, \hat{\xv}_2, \hat{\xv}_3) = \sum_{ij} \mathcal{F}^{(ij)}(\pv_1, \pv_2, \hat{\xv}_{13}, \hat{\xv}_{23}, \hat{\xv}_{3}) \left(\frac{x_{13}}{x_3} \right)^i \left (\frac{x_{23}}{x_3} \right)^j, 
\end{align}
where $\mathcal{F}^{(ij)}$ can be further decomposed into Legendre polynomials of all the angles appearing in Eq.~\eqref{F_in_wa_params}
\begin{align}
\label{F_in_legendre_poly}
    \mathcal{F}^{(ij)}(\pv_1, \pv_2, \hat{\xv}_{13}, \hat{\xv}_{23}, \hat{\xv}_{3}) &= \sum_{\ell_6+\ell_7+\ell_8 \leq i} \sum_{\ell_9+\ell_{10}+\ell_{11} \leq j} \sum_{\ell_3, \ell_4, \ell_5} \mathcal{F}^{(ij)}_{\ell_3 \ell_4 \ell_5 \ell_6 \ell_7 \ell_8 \ell_9 \ell_{10} \ell_{11}}(p_1, p_2) 
    \nn \\& \hspace{1cm} \mathcal{L}_{\ell_3}(\hat{\pv}_1 \cdot \hat{\pv}_2) \mathcal{L}_{\ell_4}(\hat{\pv}_1 \cdot \hat{\xv}_3) \mathcal{L}_{\ell_5}(\hat{\pv}_2 \cdot \hat{\xv}_3)
    \nn \\& \hspace{1cm} \mathcal{L}_{\ell_6}(\hat{\pv}_1 \cdot \hat{\xv}_{13}) \mathcal{L}_{\ell_7}(\hat{\pv}_2 \cdot \hat{\xv}_{13}) \mathcal{L}_{\ell_8}(\hat{\xv}_{13} \cdot \hat{\xv}_3)
    \nn \\& \hspace{1cm} \mathcal{L}_{\ell_9}(\hat{\pv}_1 \cdot \hat{\xv}_{23}) \mathcal{L}_{\ell_{10}}(\hat{\pv}_2 \cdot \hat{\xv}_{23}) \mathcal{L}_{\ell_{11}}(\hat{\xv}_{23} \cdot \hat{\xv}_3).
\end{align}
Once the 3PCF is expanded in Legendre multipoles, we can perform all the angular integrals in Eq.~(\ref{eq:ensamble_bisp}), arriving to the following expression for the tree-level bispectrum multipoles (see Appendix \ref{sec:wa-derivation} for derivation),
\begin{align}
\label{wa-effects}
   \langle B_\ell(&k_1, k_2, k_3) \rangle= 
    \nn \\& 
    \sum_{ij} \sum_{\ell_6+\ell_7+\ell_8 \leq i} \sum_{\ell_9+\ell_{10}+\ell_{11} \leq j} \sum_{\substack{\ell_1, \ell_2, \ell_3, \ell_4 \\  \ell_5, \ell_{12}, \ell_{13}, \ell_{14}}} i^{\ell_1 + \ell_2 - \ell_{12} - \ell_{13}} 
    \nn \\& \times \Bigg[ \frac{1}{(2\pi)^6} \int \frac{d x_3 x_3^2}{V} \int d x_{13} x_{13}^2 \int d x_{23} x_{23}^2 \int d p_1 p_1^2 \int d p_2 p_2^2
    \nn \\& \hspace{1cm} 
    j_{\ell_1}(p_1 x_{13}) j_{\ell_2}(p_2 x_{23}) j_{\ell_{12}}(k_1 x_{13}) j_{\ell_{13}}(k_2 x_{23}) \mathcal{F}^{(ij)}_{\ell_3 \ell_4 \ell_5 \ell_6 \ell_7 \ell_8 \ell_9 \ell_{10} \ell_{11}}(p_1, p_2)  \left(\frac{x_{13}}{x_3} \right)^i \left (\frac{x_{23}}{x_3} \right)^j \Bigg]
    \nn \\& \times \Bigg[
    \frac{(4\pi)^{14}}{N_{\ell_3 \ell_4 \ell_5} N_{\ell_6 \ell_7 \ell_8} N_{\ell_9 \ell_{10} \ell_{11}}} \sum_{m, m_i} \sum_{LM} (-1)^{m_3 + m_8 + m_{11} + M} \mathcal{G}_{\ell_4 \ell L}^{m_4 m M}
    \nn \\& \hspace{1cm}
    \mathcal{G}_{\ell_1 \ell_3 \ell_4 \ell_6 \ell_9}^{m_1 m_3 m_4 m_6 m_9} \mathcal{G}_{\ell_2 \ell_3 \ell_5 \ell_7 \ell_{10}}^{m_2 ,-m_3 m_5 m_7 m_{10}}
    \mathcal{G}_{\ell_1 \ell_{12} \ell_6 \ell_7 \ell_8}^{m_1 m_{12} m_6 m_7, -m_8}
    \mathcal{G}_{\ell_2 \ell_{13} \ell_9 \ell_{10} \ell_{11}}^{m_2 m_{13} m_9 m_{10}, -m_{11}} \mathcal{G}_{L \ell_5 \ell_8 \ell_{11}}^{-M m_5 m_8 m_{11}}
    \nn \\& \hspace{1cm}
    \mathcal{G}_{\ell \ell_{12} \ell_{14}}^{m m_{12}  m_{14}}Y_{\ell_{14} m_{14}}(\hat{\kv}_1) Y_{\ell_{13} m_{13}}^*(\hat{\kv}_2) \Bigg],
\end{align}
where we have defined 
\begin{equation}
    N_{\ell_1 \ell_2 \ell_3} = (2 \ell_1 + 1) (2 \ell_2 + 1) (2 \ell_3 + 1),
\end{equation}
and 
\begin{equation}
\label{G_matrices}
    \mathcal{G}_{\ell_1 \ell_2 \cdots \ell_n}^{m_1 m_2 \cdots m_n} \equiv \int d^2 \hat{n} ~ Y_{\ell_1 m_1}(\hat{\vec{n}}) Y_{\ell_2 m_2}(\hat{\vec{n}}) \cdots Y_{\ell_n m_n}(\hat{\vec{n}})\, ,
\end{equation}
which can be evaluated by repeated use of Eq.~\eqref{integration_3Y} and Eq.~\eqref{contraction_Y}.

As we can see, the final expression in Eq.~\eqref{wa-effects} includes two Hankel transforms, which can be performed analytically as shown in Appendix~\ref{sec:eval_bessel_integrations}.
In the next sections we will evaluate this expression to compute different multipoles of the bispectrum at different order in the wide-angle expansion.

\section{Numerical results}
\label{sec:num_results}

In this section, we present the numerical results of the implementation of wide-angle corrections to the bispectrum multipoles as defined in Eq.~\eqref{wa-effects}. 
We relegate all lengthy calculations to the Appendices: in particular we show how to perform the Hankel transform analytically in Appendix \ref{sec:eval_bessel_integrations} and we study the dependence to the maximum value of $\ell_3$ (which is the multipoles associated to the angle between two wave-vectors of the bispectrum $\hat{\pv}_1 \cdot \hat{\pv}_2$) in Appendix \ref{sec:convergence_test}. 

Theoretical predictions are computed using the linear power spectrum of $\Lambda$CDM cosmology, computed with CAMB code~\cite{Lewis:1999bs}, with the following parameters: $h = 0.695$ , $\Omega_c h^2 = 0.11542$, $\Omega_b h^2=0.02224$, $n_s = 0.9632$, $A_s= 2.20193 \times 10^{-9}$, $T_\mathrm{CMB} = 2.7255$. We use the quadratic fit for the quadratic bias $b_2$ \cite{Lazeyras:2015lgp} and the relation derived from excursion set formalism for the $b_{\mathcal{G}_2}$ \cite{sheth2013nonlocal, Eggemeier:2020umu}
\begin{align}
    b_{\mathcal{G}_2} &= 0.524 - 0.547~b_1 + 0.046~b_1^2,
    \\ \nonumber
    b_2 &= 0.412 - 2.143~b_1 + 0.929~b_1^2 + 0.008~b_1^3 + \frac{4}{3} b_{\mathcal{G}_2}.
\end{align}
Results are shown for two redshift bins centered at $z=1.0$ and $z=1.65$ with linear bias $b_1 = 1.46$ and $b_1 = 1.9$ respectively, following the specifications of the official \textit{Euclid} forecast \cite{Euclid:2019clj}. 

\subsection{Wide-angle effects in the monopole and dipole of the bispectrum}
\begin{figure}[t!]
    \includegraphics[width=0.99\textwidth]{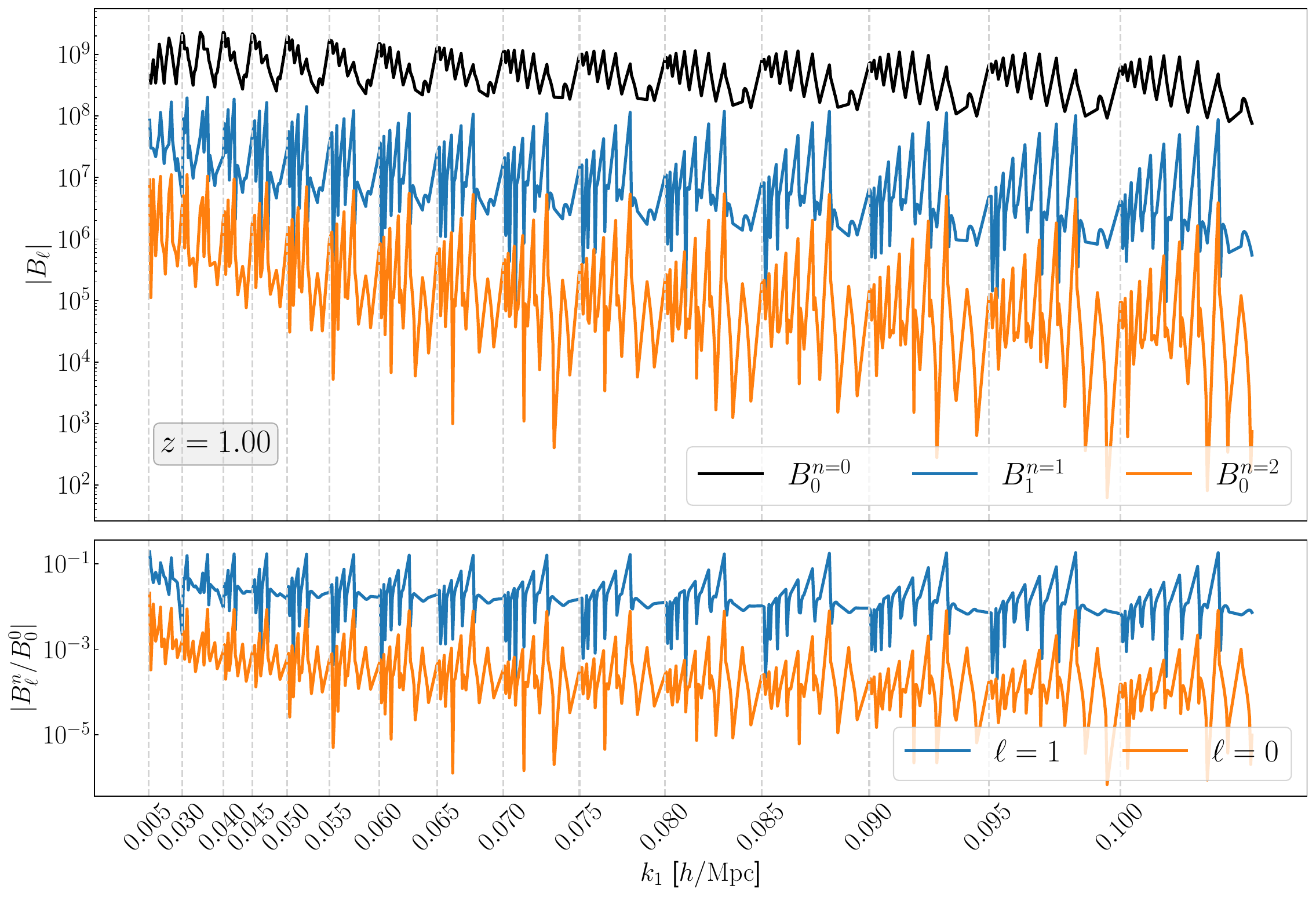}
    \centering
    \caption{The flat-sky bispectrum monopole is contrasted with the largest wide-angle corrections to the monopole and the dipole. Wide-angle corrections to the monopole are generally below $\sim 0.1 \%$ of the flat-sky monopole. Wide-angle induced dipole can reach few percents of the flat-sky bispectrum monopole. Here, the bispectrum are plotted as a function of the largest wave-number in the triangle $k_1$, for all possible triangles $(k_1, k_2, k_3)$ ordered by $k_1 \ge k_3 \ge k_2$, that satisfy the triangle conditions $k_1 \leq k_2 + k_3$ and $k_1 \ge 2 ~k_2$.}
\label{fig:the_full_triangle}
\end{figure}

\begin{figure}[h!]
    \includegraphics[width=0.99\textwidth]{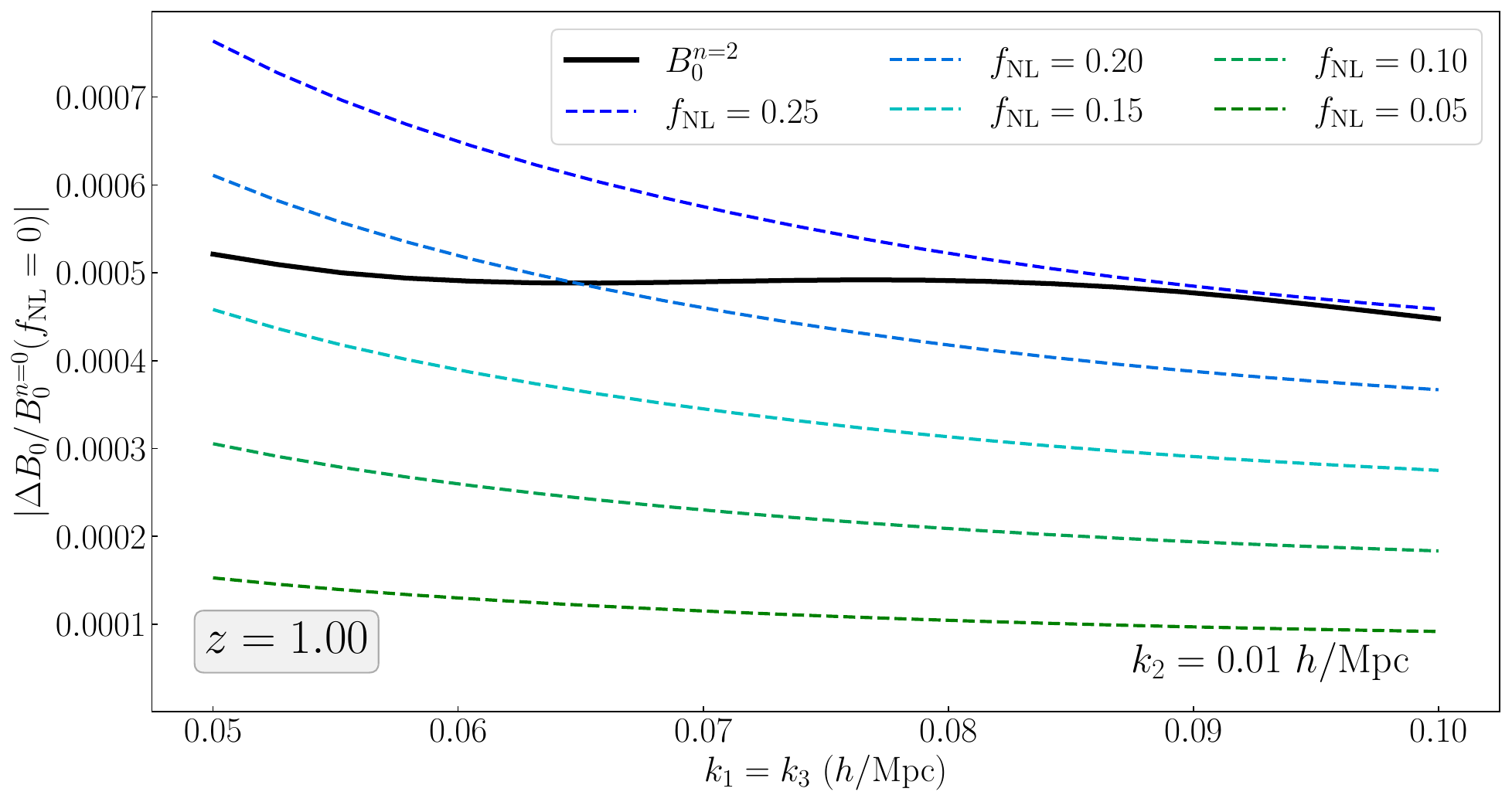}
    \centering
    \caption{Comparison of the wide-angle corrections to the monopole with the local primordial non-Gaussianity contribution. At $z=1.0$, the wide-angle contributions is comparable to a $\mathcal{O}(0.1)$ local $f_\mathrm{NL}$ signal.}
    \label{fig:the_fnl_monopole}
\end{figure}

We start by looking at the monopole of the bispectrum. As expected by symmetry, \emph{i.e.} translational invariance, the first order wide angle correction vanishes and therefore any effect start at second order, \emph{i.e.} $n \equiv i+j =2$ in Eq.~(\ref{F_in_legendre_poly}). In Figure \ref{fig:the_full_triangle} we plot, as a function of the
largest wave-number in the triangle $k_1$, all possible triangles $(k_1, k_2, k_3)$ ordered by $k_1 \ge k_3 \ge k_2$, that satisfy the triangle conditions $k_1 \leq k_2 + k_3$ and $k_1 \ge 2 ~k_2$, the amplitude of the flat-sky monopole, in black, of the leading order wide-angle correction to the monopole, in orange, and of the bispectrum dipole generated by wide-angle effects, in blue.  We see that, for most triangles, the $n=2$ correction to the flat-sky monopole is  $\sim 0.01 \%$ at $z=1$. 
On very large scales wide-angle effects approach the \% level, for a set of triangles that could be relevant to the constraints on local PNG. For this reason, in Figure \ref{fig:the_fnl_monopole} we focus on squeezed triangles, with the largest scale fixed at $k_2 = 0.01\,h$/Mpc. The dashed lines show, in units of the flat-sky monopole in a Gaussian Universe, the expected local PNG signal for different values of $f_{\rm NL}$, while the black one corresponds to the leading order WA corrections \footnote{Local PNG have been included in the theoretical prediction by modifying the perturbation theory  kernels at leading order in $f_\mathrm{NL}$ (see for example \cite{CastorinaMoradinezhad2020}).}. We find that deviations from the plane-parallel approximation roughly correspond to values of $f_{\rm NL} \sim 0.1$-$0.2$, which are close to the uncertainties  expected from a measurement of the bispectrum with future galaxy surveys \cite{Dore:2014cca,Schlegel:2019eqc,PUMA,Achucarro:2022qrl}.
It should however be kept in mind that, at fixed volume, going to higher redshifts reduces the amplitude of the WA corrections.

\begin{figure}[h!]
    \includegraphics[width=0.99\textwidth]{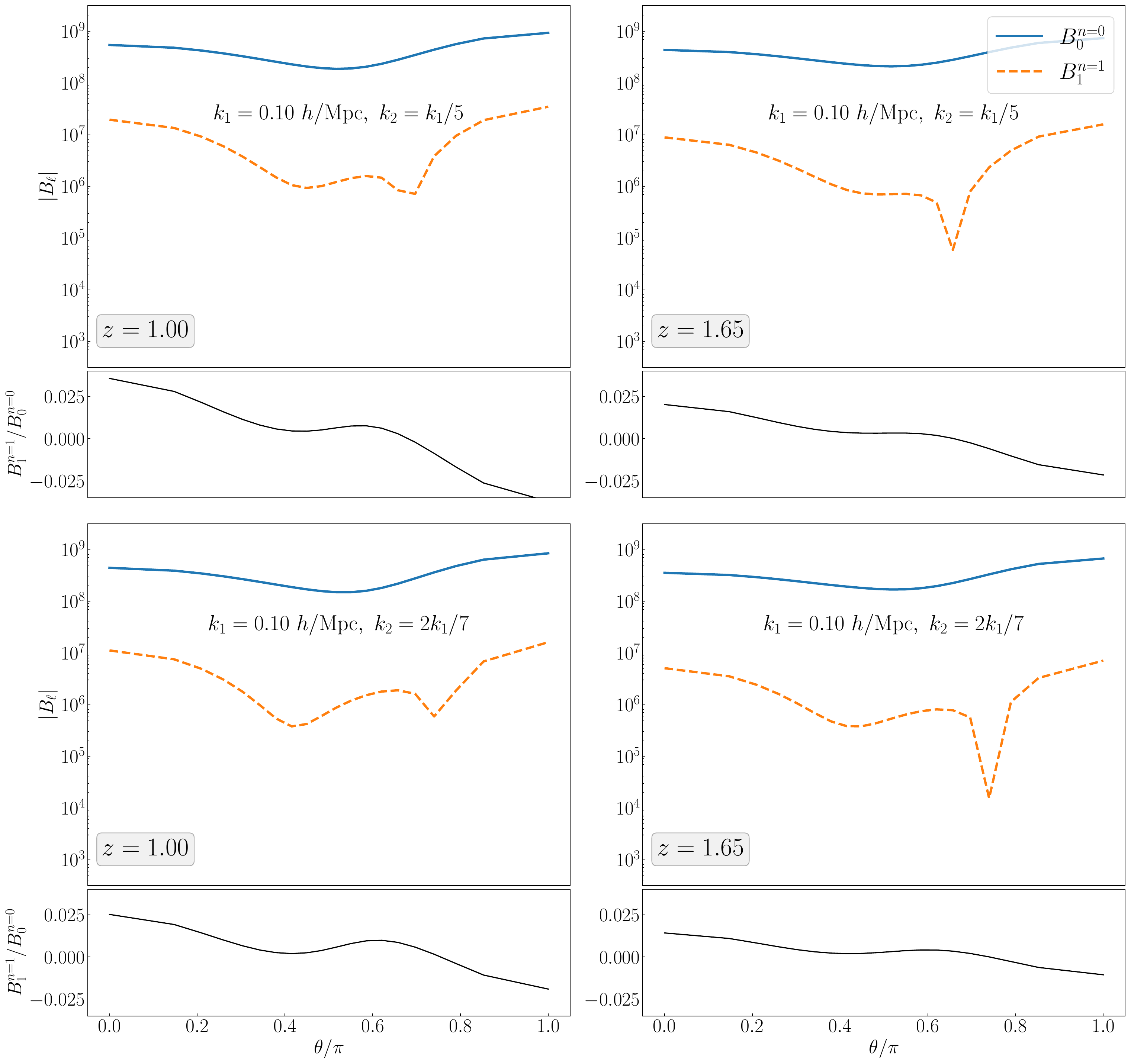}
    \centering
    \caption{Wide-angle effects induce a bispectrum dipole shown as a function of angle between the two wavevectors $\theta$, where $\cos \theta = \hat{\mathbf{k}}_1\cdot \hat{\mathbf{k}}_2$. The effects can be as large as few percents of the flat-sky bispectrum monopole.}
    \label{fig:the_dipole}
\end{figure}

Going beyond the monopole, wide-angle effects generate a non zero dipole at order $n = i+j =1$. As it can be seen from Figure \ref{fig:the_dipole}, the corrections are the order of few \%'s of the flat-sky bispectrum monopole. This holds in general, for all triangular configurations and in the redshift range $z \in [1,2]$.
This could have important implications for searches of new physics in the dipole of the bispectrum. 

\begin{figure}[h!]
    \includegraphics[width=0.99\textwidth]{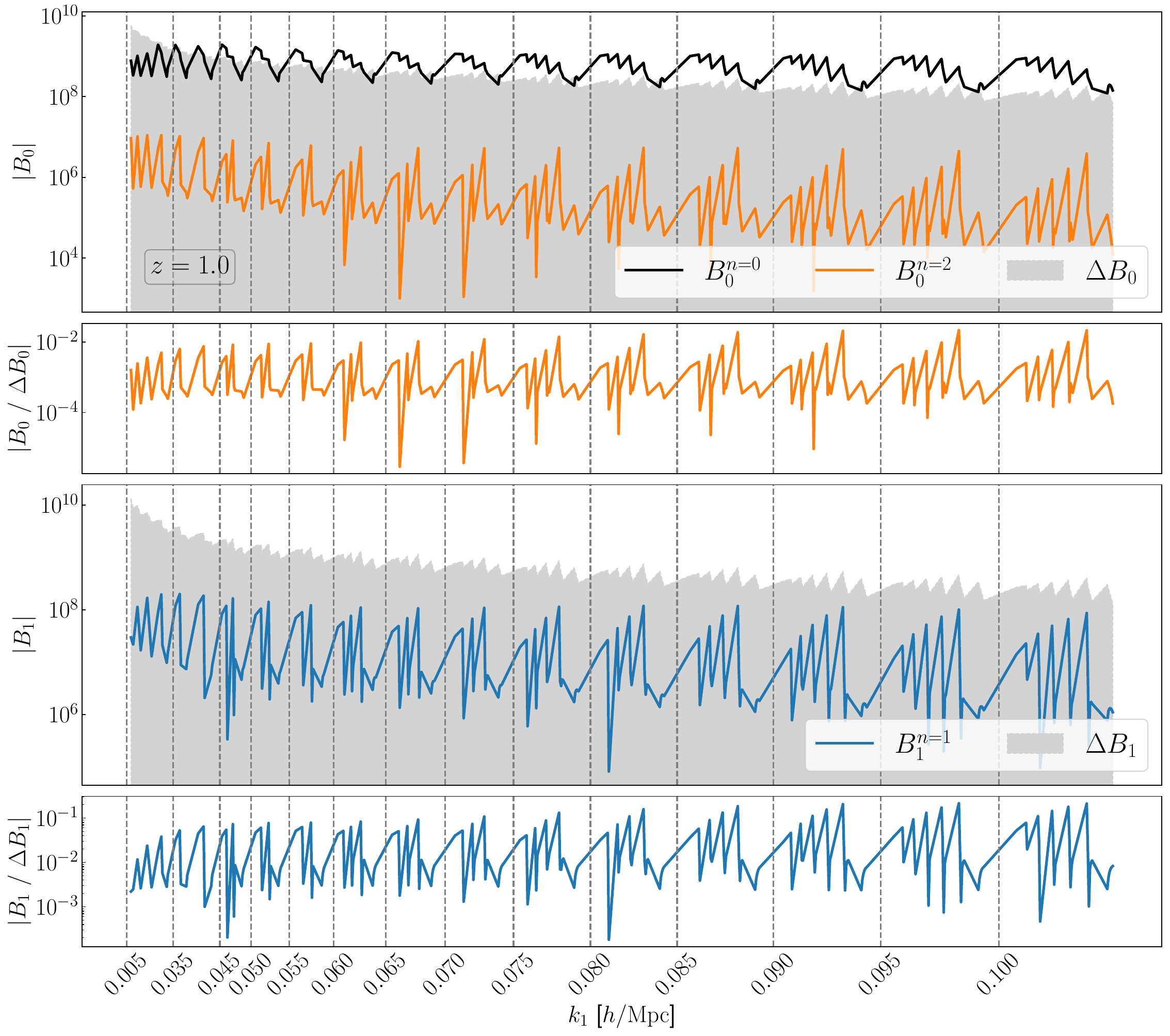}
    \centering
    \caption{Comparison of the wide-angle effects to an error budget associated with redshift bin centered at $z=1$ with total effective volume of $V_\mathrm{eff} = 8~ \mathrm{Gpc}^3 h^{-3}$ and number density $n_g = 6.86 \times 10^{-4}~h^3~\mathrm{Mpc}^{-3}$ following \cite{Euclid:2019clj}. The error is represented by a theoretical Gaussian variance assuming a flat-sky bispectrum with linear Kaiser model and Poisson shot-noise, associated with the specifications of the redshift bin. Wide-angle effects occupy at most few percents of the error of the bispectrum monopole. On the other hand, the wide-angle induced a dipole which is in the order of $\sim 10\%$ of the error budget. The bispectrum are plotted following the configurations similar to Figure \ref{fig:the_full_triangle}.}
    \label{fig:compare_with_variance}
\end{figure}

Another interesting metric to look at is how important the wide-angle effects are with respect to the error budget of the ongoing and upcoming large-volume galaxy surveys such as DESI and \textit{Euclid}. 
In Figure \ref{fig:compare_with_variance} we compare the wide-angle effects to a theoretical Gaussian variance assuming a flat-sky bispectrum with linear Kaiser model and Poisson shot-noise \footnote{See for example section 4.2 of \cite{RizzoEtal2022} for the full expression.}. Following the \textit{Euclid} forecast in \cite{Euclid:2019clj}, we consider a redshift bin centered at $z=1$ with total effective volume of $V_\mathrm{eff} = 8~ \mathrm{Gpc}^3 h^{-3}$ and number density $n_g = 6.86 \times 10^{-4}~h^3~\mathrm{Mpc}^{-3}$. We find that the wide-angle corrections to the monopole are always well below the expected cosmic variance, and can be safely neglected, unless some technique to remove sample variance is used, \emph{e.g.} multi-tracing between different samples probing the same volume. 

The signal to noise in the bispectrum dipole is instead of the order of 10\% for several triangles. Summing over all configurations indicates that the dipole could be measured with a significance of $\left(0.76\times \sqrt{\frac{V}{8\, \mathrm{Gpc}^3 h^{-3}}} \right)$ $\sigma$'s by upcoming surveys. Therefore, the wide-angle corrections to the bispectrum dipole can start being detected with surveys probing volumes larger than $50$ $\mathrm{Gpc}^3\, h^{-3}$. Note however that the presence of a window function could change this result, since the signal from other multipoles will leak into the measured dipole, similarly to what happens in the power spectrum \cite{BeutlerCastorinaZhang2018}.

We conclude this section by a qualitative discussion on the relation between WA effects and other projection effects that could change the prediction of the large scale bispectrum.
Wide-angle effects introduce corrections which scale as $(k x_3)^{-n}$ at order $n=i+j$ in the perturbative expansion. At the largest scales, relativistic effects beyond the Newtonian description in Eqs. (\ref{eq:Z1}-\ref{eq:Z2}), induces corrections which scale as $(\mathcal{H}/k)^n$.
These have been first computed in linear theory in Refs.~\cite{Yoo:2009au,Yoo:2010ni,Challinor:2011bk,Bonvin:2011bg,Jeong:2011as} and then extended to higher order in perturbation theory in Refs.~\cite{Yoo:2014sfa,Bertacca:2014dra,DiDio:2014lka,DiDio:2018zmk,DiDio:2020jvo,Magi:2022nfy}.
Therefore, we expect that these two contributions have similar amplitudes for deep surveys $x_3\sim \mathcal{H}^{-1}$, while at low redshift wide-angle effects are the main correction at large scales\footnote{ Let us note that wide-angle and relativistic effects exhibit comparable amplitudes even in the context of 2-point statistics. Furthermore, wide-angle effects are  sensitive to the definition of the line of sight. Specifically, when employing the end-point line-of-sight (as in our work), a significant enhancement of the wide-angle corrections is observed, which has already been detected in the BOSS catalog.~\cite{Gaztanaga:2015jrs,BeutlerCastorinaZhang2018}}. So far the relativistic effects has been included in the 3-dimensional Fourier bispectrum only within the flat-sky approximation\footnote{Wide-angle effects are naturally included in the angular bispectrum, see e.g.~Refs.~\cite{DiDio:2014lka,DiDio:2015bua,DiDio:2016gpd,Assassi:2017lea,DiDio:2018unb,Montandon:2022ulz}, and in the Spherical-Bessel decomposition~\cite{Bertacca:2017dzm}. However the 3-dimensional information is diluted in a large number of data points and its analysis will be hardly feasible with upcoming galaxy catalogs.}, see e.g.~\cite{Umeh:2016nuh,Jolicoeur:2017nyt,Clarkson:2018dwn,Jeong:2019igb,deWeerd:2019cae} or neglecting the convolution with a realistic window function \cite{NoorikuhaniScoccimarro2022}. Our approach can be straightforwardly extended to also incorporate relativistic corrections and, therefore, it will allow a complete and correct description of galaxy clustering at the largest scales. 

\subsection{Comparison with the Cartesian expansion formulation \cite{NoorikuhaniScoccimarro2022}}
\begin{figure}[h!]
    \includegraphics[width=0.99\textwidth]{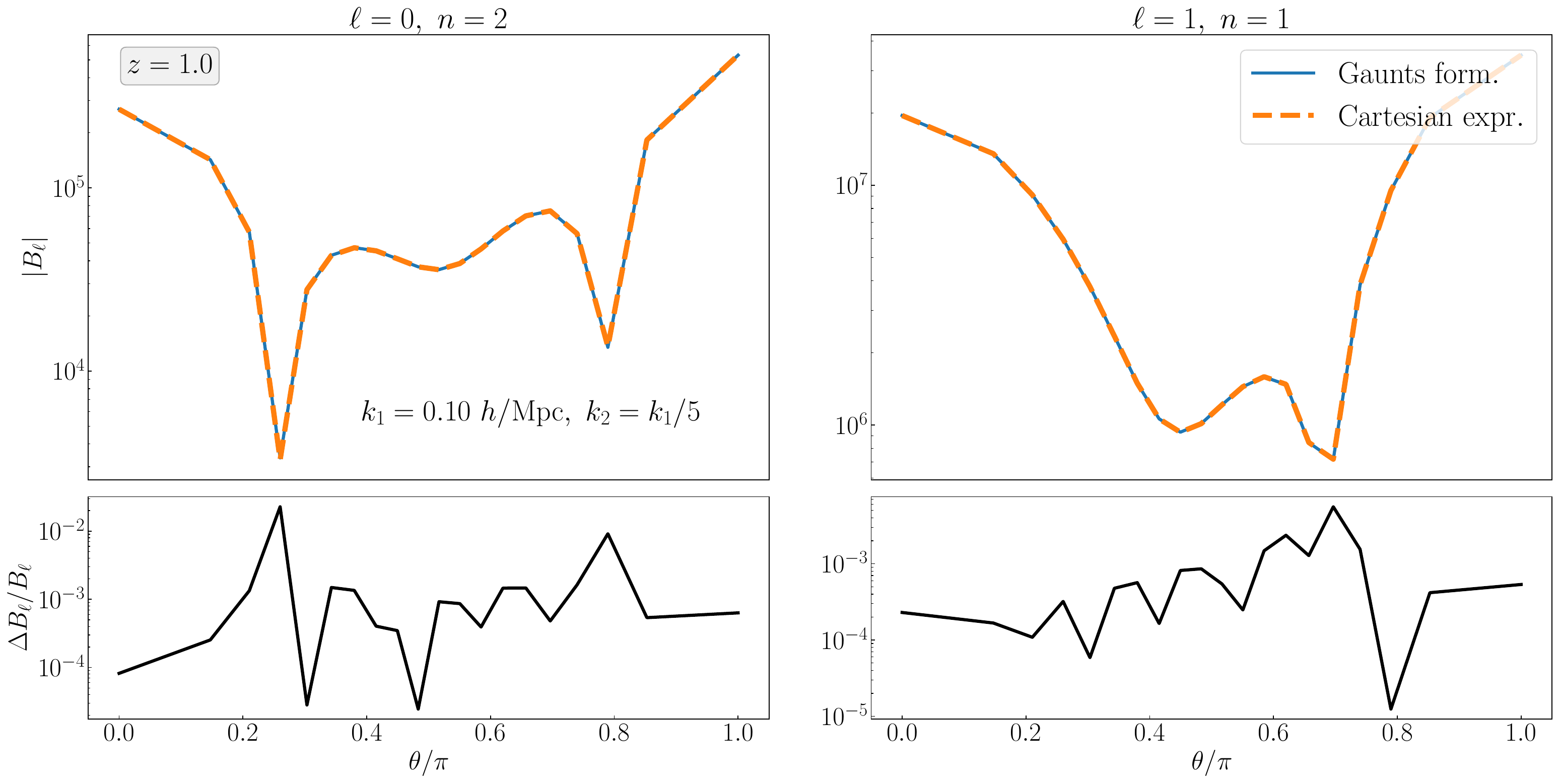}
    \centering
    \caption{Comparison of the Cartesian method developed in \cite{NoorikuhaniScoccimarro2022} and the method developed in this work with $\ell_{3,\mathrm{max}}=6$. The comparison is shown as a function of $\theta$ where $\cos \theta = \hat{\mathbf{k}}_1\cdot \hat{\mathbf{k}}_2$. The agreement is at $0.1 \%$ level for most of the triangular configurations.
    } 
    \label{fig:the_comparison}
\end{figure}

Recently, wide-angle corrections in the modeling of the bispectrum have been studied in Ref.~\cite{NoorikuhaniScoccimarro2022}. While their approach share some common features with our derivation, for instance by taking advantages of the analytical integration of the double Hankel transform, they expand the kernels in the Cartesian coordinates,
\begin{equation}
    \mathcal{F}(\pv_1, \pv_2, \hat{\xv}_1, \hat{\xv}_2, \hat{\xv}_3) = \sum_{\ell_1 + \cdots + \ell_6=n} \mathcal{F}_{\ell_1 \ell_2 \ell_3 \ell_4 \ell_5 \ell_6}(\pv_1, \pv_2, \hat{\xv}_3) x_{13,x}^{\ell_1} x_{13,y}^{\ell_2} x_{13,z}^{\ell_3} x_{23,x}^{\ell_4} x_{23,y}^{\ell_5} x_{23,z}^{\ell_6} x_3^{-n}, 
\end{equation}
where $n$ denotes the wide-angle order while $\xv_{13} = (x_{13,x}, x_{13,y}, x_{13,z})$ and similarly for $\xv_{23}$. In this way, for the most general case where we also include the window function we have
\begin{align}
    B_\ell(k_1, k_2, k_3) &= 
    (2 \ell+1) \int \frac{d^3 x_3}{V} \int d^3 x_{13} \int d^3 x_{23}~e^{-i \kv_1 \cdot \xv_{13}} e^{-i \kv_2 \cdot \xv_{23}} \int \frac{d^3 p_1}{(2\pi)^3} \frac{d^3 p_2}{(2\pi)^3} 
    \nn \\& \hspace{0.5cm} 
    e^{i \pv_1 \cdot \xv_{13}} e^{i \pv_2 \cdot \xv_{23}}\mathcal{F}(\pv_1, \pv_2, \hat{\xv}_1, \hat{\xv}_2, \hat{\xv}_3) W(\xv_1) W(\xv_2) W(\xv_3) \mathcal{L}_\ell (\kv_1 \cdot \xv_3)
    \nn \\&=
    (2 \ell+1) \int \frac{d^2 \hat{x}_3}{V} \int d^3 x_{13} \int d^3 x_{23}~e^{-i \kv_1 \cdot \xv_{13}} e^{-i \kv_2 \cdot \xv_{23}} \int \frac{d^3 p_1}{(2\pi)^3} \frac{d^3 p_2}{(2\pi)^3} 
    \nn \\& \hspace{0.5cm} 
    e^{i \pv_1 \cdot \xv_{13}} e^{i \pv_2 \cdot \xv_{23}} \sum_{\ell_1 + \cdots + \ell_6=n} \mathcal{F}_{\ell_1 \ell_2 \ell_3 \ell_4 \ell_5 \ell_6}(\pv_1, \pv_2, \hat{\xv}_3) x_{13,x}^{\ell_1} x_{13,y}^{\ell_2} x_{13,z}^{\ell_3} x_{23,x}^{\ell_4} x_{23,y}^{\ell_5} x_{23,z}^{\ell_6} 
    \nn \\& \hspace{0.5cm}
    \sum_{J_1, J_2, J_3} Q^{(n)}_{J_1 J_2 J_3}(x_{13}, x_{23}) \mathcal{L}_{J_1}(\hat \xv_{13} \cdot \hat \xv_3) \mathcal{L}_{J_2}(\hat \xv_{23} \cdot \hat \xv_3) \mathcal{L}_{J_3}(\hat \xv_{13} \cdot \hat \xv_{23}) \mathcal{L}_\ell (\kv_1 \cdot \xv_3),
\end{align}
where we have defined the 3PCF of the window as follows
\begin{align}
\label{3pcf_cartesian}
    Q^{(n)}(\xv_{13}, \xv_{23}, \hat \xv_3) &\equiv \int  \frac{dx_3}{V} x^{2-n}_3 ~W(\xv_1) W(\xv_2) W(\xv_3) 
    \nn \\&=
    \sum_{J_1, J_2, J_3} Q^{(n)}_{J_1 J_2 J_3}(x_{13}, x_{23}) \mathcal{L}_{J_1}(\hat \xv_{13} \cdot \hat \xv_3) \mathcal{L}_{J_2}(\hat \xv_{23} \cdot \hat \xv_3) \mathcal{L}_{J_3}(\hat \xv_{13} \cdot \hat \xv_{23}).
\end{align}
In the case of uniform window function $W(\xv) = 1$,
\begin{align}
Q_{J_1 J_2 J_3}^{(n)}(x_{13}, x_{23}) = \int \frac{dx_3}{V} x_3^{2-n} \delta_{J_1 0} \delta_{J_2 0} \delta_{J_3 0},
\end{align}
and thus the Cartesian coordinates of $\xv_{13}$ and $\xv_{23}$ only appear as some polynomial functions. Denoting $\kv_1 = (k_1^x, k_1^y, k_1^z)$ (and similarly for $\kv_2$) we can replace 
\begin{equation}
x^{\ell_1}_{13,x} \rightarrow (-i)^{\ell_1} \partial_{k_1^x}^{\ell_1}e^{-i \kv_1 \cdot \xv_{13}},
\end{equation}
and similarly for the others, giving
\begin{align}
    B_\ell(&k_1, k_2, k_3)  
    \nn \\&= 
    (2\ell+1) (-i)^{-n} \sum_{\ell_1 + \cdots + \ell_6=n} \partial_{k_1^x}^{\ell_1} \partial_{k_1^y}^{\ell_2} \partial_{k_1^z}^{\ell_3} \partial_{k_2^x}^{\ell_4} \partial_{k_2^y}^{\ell_5} \partial_{k_2^z}^{\ell_6}
    \nn  \\& \hspace{0.5cm} \times
    \int \frac{d^3 x_3}{V} \int d^3 x_{13} \int d^3 x_{23}~e^{-i \kv_1 \cdot \xv_{13}} e^{-i \kv_2 \cdot \xv_{23}} \int \frac{d^3 p_1}{(2\pi)^3} \frac{d^3 p_2}{(2\pi)^3} e^{i \pv_1 \cdot \xv_{13}} e^{i \pv_2 \cdot \xv_{23}}
    \nn \\& \hspace{1cm}  \mathcal{F}_{\ell_1 \ell_2 \ell_3 \ell_4 \ell_5 \ell_6}(\pv_1, \pv_2, \hat{\xv}_3)  \mathcal{L}_\ell (\kv_1 \cdot \xv_3) x_3^{-n}
    \nn \\&= (2\ell+1) (-i)^{-n} \sum_{\ell_1 + \cdots + \ell_6=n} \partial_{k_1^x}^{\ell_1} \partial_{k_1^y}^{\ell_2} \partial_{k_1^z}^{\ell_3} \partial_{k_2^x}^{\ell_4} \partial_{k_2^y}^{\ell_5} \partial_{k_2^z}^{\ell_6} 
    \nn  \\& \hspace{0.5cm} \times
    \int \frac{d^3 x_3}{(2\pi)^6 V} \mathcal{F}_{\ell_1 \ell_2 \ell_3 \ell_4 \ell_5 \ell_6}(\kv_1, \kv_2, \hat{\xv}_3)  \mathcal{L}_\ell (\kv_1 \cdot \xv_3) x_3^{-n}.
\end{align}
The main advantage of using Cartesian coordinates relies on having only finite sums. Therefore we can use it first to cross-check our results and then also to study the convergence and the cutoff of the sum over $\ell_3$ in Eq.~\eqref{wa-effects}.
In Figure \ref{fig:the_comparison}, we show that both Cartesian method and the method developed in this work agree at $0.1 \%$ level for most of the triangular configurations.
We have also checked that our results agree numerically with Ref.~\cite{NoorikuhaniScoccimarro2022} whenever a comparison was possible. 

The main drawback of the Cartesian coordinates method presented above is that it is not particularly well suited to include any realistic window function. 
Actually, the Cartesian method relies on the assumption that the Cartesian coordinates of $\xv_{13}$ and $\xv_{23}$ appear as some polynomial functions, which only true for uniform window function $W(\xv)=1$. The Cartesian approach can still be generalized to include more realistic geometry, but at the price of a much larger computational complexity. 

On the other hand, the method presented in this work, as shown in Appendix \ref{sec:wa-derivation}, can be trivially extended to include arbitrary window functions. The main difference with the formulation presented in the main text is that now the Hankel transforms in Eq.~(\ref{wa-effects}) have to be evaluated numerically via FFTs (e.g. with 2D-FFTLog \cite{Fang2DFFTLog2020}).

\section{Conclusions and outlook}
\label{sec:conclusion}
In this work we laid out a formalism to include wide-angle effects in the bispectrum multipoles,  matching the theoretical model to the most commonly used bispectum estimator. We found that, at $z=1$, wide-angle effects on the monopole are, for most triangular configurations, a $\sim 0.1 \%$ correction to the flat-sky monopole. The wide-angle contributions to the monopole however can become comparable to a $\mathcal{O}(0.1)$ local $f_\mathrm{NL}$ signal in the squeezed limit. Wide-angle effects also can induce a dipole that can reach few percents of the flat-sky bispectrum monopole. In general this effect is less pronounced at higher redshift and at smaller-scale. 

We also consider the prospect of detecting these effects for the ongoing and upcoming galaxy redshift surveys such as DESI and \textit{Euclid}. We found that for a typical reshift-bin of total volume $V_\mathrm{eff} = 8~ \mathrm{Gpc}^3 h^{-3}$, wide-angle effects in the monopole can be safely neglected. The bispectrum dipole could instead be detected by surveys probing volumes larger than $50$ $\mathrm{Gpc}^3\, h^{-3}$.
Note that these numbers will change accordingly if the effective redshift of the sample changes.

Future direction which we have not explored in this work include a numerically efficient implementation of wide-angle effects in the presence of a window function, and the generalization of our method to include relativistic effects. We plan to return to these interesting problems in a future work.

\acknowledgments
We would like to thank Matteo Biagetti, Emiliano Sefusatti and Oliver Philcox for valuable discussions. K.P. is partially supported by the INFN INDARK PD51 grant.

\appendix

\section{Derivation of the bispectrum wide-angle formulation}
\label{sec:wa-derivation}

In this appendix we explicitly compute the ensemble average of the bispectrum estimator beyond the plane-parallel approximation.
From the definition of the bispectrum multipoles including the window functions $W(\xv)$ we have 
\begin{align}
 \langle   \hat B_\ell(&k_1, k_2, k_3) \rangle
    \nn \\&= (2 \ell+1) \int  \frac{d^3 x_3}{V} \int d^3 x_{13} \int d^3 x_{23} ~ \langle \delta_g(\xv_1) \delta_g(\xv_2) \delta_g(\xv_3) \rangle 
    \nn \\& \hspace{0.5cm} \times W(\xv_1) W(\xv_2) W(\xv_3) e^{-i \kv_1 \cdot \xv_{13}} e^{-i \kv_2 \cdot \xv_{23}} \mathcal{L}_\ell(\hat{\kv}_1 \cdot \hat{\xv}_3)
    \nn \\&= (2 \ell+1) \int \frac{d^3 x_3}{V} \int d^3 x_{13} \int d^3 x_{23}~e^{-i \kv_1 \cdot \xv_{13}} e^{-i \kv_2 \cdot \xv_{23}}
    \nn \\& \hspace{0.5cm} \times \int \frac{d^3 p_1}{(2\pi)^3} \frac{d^3 p_2}{(2\pi)^3} e^{i \pv_1 \cdot \xv_{13}} e^{i \pv_2 \cdot \xv_{23}} \mathcal{F}(\pv_1, \pv_2, \hat{\xv}_1, \hat{\xv}_2, \hat{\xv}_3) W(\xv_1) W(\xv_2) W(\xv_3) \mathcal{L}_\ell (\hat{\kv}_1 \cdot \hat{\xv}_3),
\end{align}
where we have used Eq.~\eqref{3pcf_beyond_pp} and Eq.~\eqref{F_kernels}. Taking $\hat{\xv}_3$ as the line of sight, we can expand the $\mathcal{F}$ kernels as in Eq.~\eqref{F_in_legendre_poly}. 
\begin{align}
  \langle \hat  B_\ell(&k_1, k_2, k_3) \rangle
    \nn \\&= (2 \ell+1) \int \frac{d^3 x_3}{V} \int d^3 x_{13} \int d^3 x_{23}~e^{-i \kv_1 \cdot \xv_{13}} e^{-i \kv_2 \cdot \xv_{23}}
    \nn \\& \hspace{0.5cm} \times 
    \int \frac{d^3 p_1}{(2\pi)^3} \frac{d^3 p_2}{(2\pi)^3} e^{i \pv_1 \cdot \xv_{13}} e^{i \pv_2 \cdot \xv_{23}} \nn \\& \hspace{0.5cm} \times 
    \sum_{ij} \sum_{\ell_6+\ell_7+\ell_8 \leq i} \sum_{\ell_9+\ell_{10}+\ell_{11} \leq j} \sum_{\ell_3, \ell_4, \ell_5} \mathcal{F}^{(ij)}_{\ell_3 \ell_4 \ell_5 \ell_6 \ell_7 \ell_8 \ell_9 \ell_{10} \ell_{11}}(p_1, p_2)  \left(\frac{x_{13}}{x_3} \right)^i \left (\frac{x_{23}}{x_3} \right)^j
    \nn \\& \hspace{1cm} \mathcal{L}_{\ell_3}(\hat{\pv}_1 \cdot \hat{\pv}_2) \mathcal{L}_{\ell_4}(\hat{\pv}_1 \cdot \hat{\xv}_3) \mathcal{L}_{\ell_5}(\hat{\pv}_2 \cdot \hat{\xv}_3)
    \nn \\& \hspace{1cm} \mathcal{L}_{\ell_6}(\hat{\pv}_1 \cdot \hat{\xv}_{13}) \mathcal{L}_{\ell_7}(\hat{\pv}_2 \cdot \hat{\xv}_{13}) \mathcal{L}_{\ell_8}(\hat{\xv}_{13} \cdot \hat{\xv}_3)
    \nn \\& \hspace{1cm} \mathcal{L}_{\ell_9}(\hat{\pv}_1 \cdot \hat{\xv}_{23}) \mathcal{L}_{\ell_{10}}(\hat{\pv}_2 \cdot \hat{\xv}_{23}) \mathcal{L}_{\ell_{11}}(\hat{\xv}_{23} \cdot \hat{\xv}_3) \mathcal{L}_\ell(\hat{\kv}_1 \cdot \hat{\xv}_3) W(\xv_1) W(\xv_2) W(\xv_3),
\end{align}
which can be further organized in terms of Legendre polynomials by expanding the plane wave using the Rayleigh expansion identity Eq.~\eqref{plane_wave_exp}
\begin{align}
\langle \hat B_\ell(&k_1, k_2, k_3)  \rangle
    \nn \\&= \frac{2 \ell+1}{(2\pi)^6} \int \frac{d x_3 x_3^2}{V} \int d x_{13} x_{13}^2 \int d x_{23} x_{23}^2 \int d p_1 p_1^2 \int d p_2 p_2^2
    \nn \\& \hspace{0.5cm} \times 
    \sum_{ij} \sum_{\ell_6+\ell_7+\ell_8 \leq i} \sum_{\ell_9+\ell_{10}+\ell_{11} \leq j} \sum_{\ell_1, \ell_2, \ell_3, \ell_4, \ell_5, \ell_{12}, \ell_{13}} 
    \nn \\& \hspace{1 cm} 
    i^{\ell_1 + \ell_2 - \ell_{12} - \ell_{13}} (2\ell_1+1) (2\ell_2+1) (2\ell_{12}+1)(2\ell_{13}+1) 
    \nn \\& \hspace{1cm}
    j_{\ell_1}(p_1 x_{13}) j_{\ell_2}(p_2 x_{23}) j_{\ell_{12}}(k_1 x_{13}) j_{\ell_{13}}(k_2 x_{23}) \mathcal{F}^{(ij)}_{\ell_3 \ell_4 \ell_5 \ell_6 \ell_7 \ell_8 \ell_9 \ell_{10} \ell_{11}}(p_1, p_2)  \left(\frac{x_{13}}{x_3} \right)^i \left (\frac{x_{23}}{x_3} \right)^j
    \nn \\& \hspace{1cm}
    \int d^2 \hat{x}_3 \int d^2 \hat{x}_{13} \int d^2 \hat{x}_{23} \int d^2 \hat{p}_{1} \int d^2 \hat{p}_{2}
    \nn \\& \hspace{1cm}
    \mathcal{L}_{\ell_1}(\hat{\pv}_1 \cdot \hat{\xv}_{13}) \mathcal{L}_{\ell_2}(\hat{\pv}_2 \cdot \hat{\xv}_{23})
    \mathcal{L}_{\ell_{12}}(\hat{\kv}_1 \cdot \hat{\xv}_{13})
    \mathcal{L}_{\ell_{13}}(\hat{\kv}_2 \cdot \hat{\xv}_{23})
    \nn \\& \hspace{1cm} \mathcal{L}_{\ell_3}(\hat{\pv}_1 \cdot \hat{\pv}_2) \mathcal{L}_{\ell_4}(\hat{\pv}_1 \cdot \hat{\xv}_3) \mathcal{L}_{\ell_5}(\hat{\pv}_2 \cdot \hat{\xv}_3)
    \nn \\& \hspace{1cm} \mathcal{L}_{\ell_6}(\hat{\pv}_1 \cdot \hat{\xv}_{13}) \mathcal{L}_{\ell_7}(\hat{\pv}_2 \cdot \hat{\xv}_{13}) \mathcal{L}_{\ell_8}(\hat{\xv}_{13} \cdot \hat{\xv}_3)
    \nn \\& \hspace{1cm} \mathcal{L}_{\ell_9}(\hat{\pv}_1 \cdot \hat{\xv}_{23}) \mathcal{L}_{\ell_{10}}(\hat{\pv}_2 \cdot \hat{\xv}_{23}) \mathcal{L}_{\ell_{11}}(\hat{\xv}_{23} \cdot \hat{\xv}_3) \mathcal{L}_\ell(\hat{\kv}_1 \cdot \hat{\xv}_3) W(\xv_1) W(\xv_2) W(\xv_3)\,.
\end{align}
Let us then define the 3PCF of the window as follows
\begin{align}
\label{3pcf_gaunt_form}
    Q^{(n)}(\xv_{13}, \xv_{23}, \hat{\xv}_3) &\equiv \int  \frac{dx_3}{V} x_3^{2-n} W(\xv_1) W(\xv_2) W(\xv_3)
    \nn \\&=
    \sum_{J_1, J_2, J_3} Q^{(n)}_{J_1 J_2 J_3}(x_{13}, x_{23}) \mathcal{L}_{J_1}(\hat{\xv}_{13} \cdot \hat{\xv}_{23}) \mathcal{L}_{J_2}(\hat{\xv}_{13} \cdot \hat{\xv}_{3}) \mathcal{L}_{J_3}(\hat{\xv}_{23} \cdot \hat{\xv}_{3}),
\end{align}
so that now
\begin{align}
 &\langle \hat  B_\ell(k_1, k_2, k_3) \rangle
    \nn \\&= \frac{2 \ell+1}{(2\pi)^6} \int d x_{13} x_{13}^2 \int d x_{23} x_{23}^2 \int d p_1 p_1^2 \int d p_2 p_2^2
    \nn \\& \hspace{0.5cm} \times 
    \sum_{ij} \sum_{\ell_6+\ell_7+\ell_8 \leq i} \sum_{\ell_9+\ell_{10}+\ell_{11} \leq j} \sum_{\ell_1, \ell_2, \ell_3, \ell_4, \ell_5, \ell_{12}, \ell_{13}} 
    \nn \\& \hspace{1 cm} 
    i^{\ell_1 + \ell_2 - \ell_{12} - \ell_{13}} (2\ell_1+1) (2\ell_2+1) (2\ell_{12}+1)(2\ell_{13}+1) 
    \nn \\& \hspace{1cm}
    j_{\ell_1}(p_1 x_{13}) j_{\ell_2}(p_2 x_{23}) j_{\ell_{12}}(k_1 x_{13}) j_{\ell_{13}}(k_2 x_{23}) \mathcal{F}^{(ij)}_{\ell_3 \ell_4 \ell_5 \ell_6 \ell_7 \ell_8 \ell_9 \ell_{10} \ell_{11}}(p_1, p_2)  x_{13}^i x_{23}^j
    \nn \\& \hspace{1cm}
    \int d^2 \hat{x}_3 \int d^2 \hat{x}_{13} \int d^2 \hat{x}_{23} \int d^2 \hat{p}_{1} \int d^2 \hat{p}_{2}
    \nn \\& \hspace{1cm}
    \mathcal{L}_{\ell_1}(\hat{\pv}_1 \cdot \hat{\xv}_{13}) \mathcal{L}_{\ell_2}(\hat{\pv}_2 \cdot \hat{\xv}_{23})
    \mathcal{L}_{\ell_{12}}(\hat{\kv}_1 \cdot \hat{\xv}_{13})
    \mathcal{L}_{\ell_{13}}(\hat{\kv}_2 \cdot \hat{\xv}_{23})
    \nn \\& \hspace{1cm} \mathcal{L}_{\ell_3}(\hat{\pv}_1 \cdot \hat{\pv}_2) \mathcal{L}_{\ell_4}(\hat{\pv}_1 \cdot \hat{\xv}_3) \mathcal{L}_{\ell_5}(\hat{\pv}_2 \cdot \hat{\xv}_3)
    \nn \\& \hspace{1cm} \mathcal{L}_{\ell_6}(\hat{\pv}_1 \cdot \hat{\xv}_{13}) \mathcal{L}_{\ell_7}(\hat{\pv}_2 \cdot \hat{\xv}_{13}) \mathcal{L}_{\ell_8}(\hat{\xv}_{13} \cdot \hat{\xv}_3)
    \nn \\& \hspace{1cm} \mathcal{L}_{\ell_9}(\hat{\pv}_1 \cdot \hat{\xv}_{23}) \mathcal{L}_{\ell_{10}}(\hat{\pv}_2 \cdot \hat{\xv}_{23}) \mathcal{L}_{\ell_{11}}(\hat{\xv}_{23} \cdot \hat{\xv}_3) \mathcal{L}_\ell(\hat{\kv}_1 \cdot \hat{\xv}_3)
    \nn \\& \hspace{1cm} \sum_{J_1, J_2, J_3} Q_{J_1 J_2 J_3}^{(i+j)}(x_{13}, x_{23}) \mathcal{L}_{J_1}(\hat{\xv}_{13} \cdot \hat{\xv}_{23}) \mathcal{L}_{J_2}(\hat{\xv}_{13} \cdot \hat{\xv}_{3})
    \mathcal{L}_{J_3}(\hat{\xv}_{23} \cdot \hat{\xv}_{3})
    \nn \\&= \frac{2 \ell+1}{(2\pi)^6} \int d x_{13} x_{13}^2 \int d x_{23} x_{23}^2 \int d p_1 p_1^2 \int d p_2 p_2^2
    \nn \\& \hspace{0.5cm} \times 
    \sum_{ij} \sum_{\ell_6+\ell_7+\ell_8 \leq i} \sum_{\ell_9+\ell_{10}+\ell_{11} \leq j} \sum_{\ell_1, \ell_2, \ell_3, \ell_4, \ell_5, \ell_{12}, \ell_{13}} \sum_{J_1, J_2, J_3}
    \nn \\& \hspace{1 cm} 
    i^{\ell_1 + \ell_2 - \ell_{12} - \ell_{13}} (2\ell_1+1) (2\ell_2+1) (2\ell_{12}+1)(2\ell_{13}+1) j_{\ell_1}(p_1 x_{13}) j_{\ell_2}(p_2 x_{23}) 
    \nn \\& \hspace{1cm}
    j_{\ell_{12}}(k_1 x_{13}) j_{\ell_{13}}(k_2 x_{23}) \mathcal{F}^{(ij)}_{\ell_3 \ell_4 \ell_5 \ell_6 \ell_7 \ell_8 \ell_9 \ell_{10} \ell_{11}}(p_1, p_2) Q_{J_1 J_2 J_3}^{(i+j)}(x_{13}, x_{23})  x_{13}^i x_{23}^j
    \nn \\& \hspace{1cm}
    \int d^2 \hat{x}_3 \int d^2 \hat{x}_{13} \int d^2 \hat{x}_{23} \int d^2 \hat{p}_{1} \int d^2 \hat{p}_{2}
    \nn \\& \hspace{1cm}
    \sum_{\ell'_8, \ell'_{11}} 
    (2\ell'_8+1)(2\ell'_{11}+1)
    \begin{pmatrix}
    J_2 & \ell_8 & \ell'_8 \\
    0 & 0 & 0
    \end{pmatrix}^2
    \begin{pmatrix}
    J_3 & \ell_{11} & \ell'_{11} \\
    0 & 0 & 0
    \end{pmatrix}^2
    \nn \\& \hspace{1cm}
    \mathcal{L}_{\ell_1}(\hat{\pv}_1 \cdot \hat{\xv}_{13}) \mathcal{L}_{\ell_2}(\hat{\pv}_2 \cdot \hat{\xv}_{23})
    \mathcal{L}_{\ell_{12}}(\hat{\kv}_1 \cdot \hat{\xv}_{13})
    \mathcal{L}_{\ell_{13}}(\hat{\kv}_2 \cdot \hat{\xv}_{23})
    \nn \\& \hspace{1cm} \mathcal{L}_{\ell_3}(\hat{\pv}_1 \cdot \hat{\pv}_2) \mathcal{L}_{\ell_4}(\hat{\pv}_1 \cdot \hat{\xv}_3) \mathcal{L}_{\ell_5}(\hat{\pv}_2 \cdot \hat{\xv}_3)
    \nn \\& \hspace{1cm} \mathcal{L}_{\ell_6}(\hat{\pv}_1 \cdot \hat{\xv}_{13}) \mathcal{L}_{\ell_7}(\hat{\pv}_2 \cdot \hat{\xv}_{13}) \mathcal{L}_{\ell'_8}(\hat{\xv}_{13} \cdot \hat{\xv}_3)
    \nn \\& \hspace{1cm} \mathcal{L}_{\ell_9}(\hat{\pv}_1 \cdot \hat{\xv}_{23}) \mathcal{L}_{\ell_{10}}(\hat{\pv}_2 \cdot \hat{\xv}_{23}) \mathcal{L}_{\ell'_{11}}(\hat{\xv}_{23} \cdot \hat{\xv}_3) \mathcal{L}_\ell(\hat{\kv}_1 \cdot \hat{\xv}_3) 
    \mathcal{L}_{J_1}(\hat{\xv}_{13} \cdot \hat{\xv}_{23}),
\end{align}    
where we have used the product of Legendre polynomials identity Eq.~\eqref{contraction_legendre}.
Expanding the Legendre polynomials in terms of spherical harmonics, Eq.~\eqref{addition_spherical_harmonics}, we can collect the angular integrations in terms of the matrices defined in Eq.~\eqref{G_matrices}
\begin{align}
\label{final-step-derivation}
    &\langle \hat B_\ell(k_1, k_2, k_3) \rangle
    \nn \\&= \frac{1}{(2\pi)^6}  \int d x_{13} x_{13}^2 \int d x_{23} x_{23}^2 \int d p_1 p_1^2 \int d p_2 p_2^2
    \nn \\& \hspace{0.5cm} \times 
    \sum_{ij} \sum_{\ell_6+\ell_7+\ell_8 \leq i} \sum_{\ell_9+\ell_{10}+\ell_{11} \leq j} \sum_{\ell_1, \ell_2, \ell_3, \ell_4, \ell_5, \ell_{12}, \ell_{13}} \sum_{J_1, J_2, J_3} i^{\ell_1 + \ell_2 - \ell_{12} - \ell_{13}} j_{\ell_1}(p_1 x_{13}) j_{\ell_2}(p_2 x_{23})
    \nn \\& \hspace{1cm} 
    j_{\ell_{12}}(k_1 x_{13}) j_{\ell_{13}}(k_2 x_{23}) \mathcal{F}^{(ij)}_{\ell_3 \ell_4 \ell_5 \ell_6 \ell_7 \ell_8 \ell_9 \ell_{10} \ell_{11}}(p_1, p_2) Q_{J_1 J_2 J_3}^{(i+j)}(x_{13}, x_{23}) x_{13}^i x_{23}^j
    \nn \\& \hspace{1cm}
    \int d^2 \hat{x}_3 \int d^2 \hat{x}_{13} \int d^2 \hat{x}_{23} \int d^2 \hat{p}_{1} \int d^2 \hat{p}_{2} 
    \nn \\& \hspace{1cm}
    \sum_{\ell'_8, \ell'_{11}} 
    \begin{pmatrix}
    J_2 & \ell_8 & \ell'_8 \\
    0 & 0 & 0
    \end{pmatrix}^2
    \begin{pmatrix}
    J_3 & \ell_{11} & \ell'_{11} \\
    0 & 0 & 0
    \end{pmatrix}^2
    \frac{(4\pi)^{15}}{N_{\ell_3 \ell_4 \ell_5} N_{\ell_6 \ell_7 \ell_9} N_{\ell_{10} J_1 0}} 
    \nn \\& \hspace{1cm}
    \sum_{m, m_i, m'_i, j_i} Y_{\ell_1 m_1}^*(\hat{\pv}_1) Y_{\ell_1 m_1}(\hat{\xv}_{13}) Y_{\ell_2 m_2}^*(\hat{\pv}_2) Y_{\ell_2 m_2}(\hat{\xv}_{23})
    \nn \\& \hspace{1cm}
    Y_{\ell_{12} m_{12}}^*(\hat{\kv}_1) Y_{\ell_{12} m_{12}}(\hat{\xv}_{13}) Y_{\ell_{13} m_{13}}^*(\hat{\kv}_2) Y_{\ell_{13} m_{13}}(\hat{\xv}_{23})
    \nn \\& \hspace{1cm}
    Y_{\ell_3 m_3}^*(\hat{\pv}_1) Y_{\ell_3 m_3}(\hat{\pv}_2) Y_{\ell_4 m_4}^*(\hat{\pv}_1) Y_{\ell_4 m_4}(\hat{\xv}_3)
    Y_{\ell_5 m_5}^*(\hat{\pv}_2) Y_{\ell_5 m_5}(\hat{\xv}_3)
    \nn \\& \hspace{1cm}
    Y_{\ell_6 m_6}^*(\hat{\pv}_1) Y_{\ell_6 m_6}(\hat{\xv}_{13})
    Y_{\ell_7 m_7}^*(\hat{\pv}_2) Y_{\ell_7 m_7}(\hat{\xv}_{13})
    Y_{\ell'_8 m'_8}^*(\hat{\xv}_{13}) Y_{\ell'_8 m'_8}(\hat{\xv}_{3})
    \nn \\& \hspace{1cm}
    Y_{\ell_9 m_9}^*(\hat{\pv}_1) Y_{\ell_9 m_9}(\hat{\xv}_{23})
    Y_{\ell_{10} m_{10}}^*(\hat{\pv}_2) Y_{\ell_{10} m_{10}}(\hat{\xv}_{23})
    Y_{\ell'_{11} m'_{11}}^*(\hat{\xv}_{23}) Y_{\ell'_{11} m'_{11}}(\hat{\xv}_{3})
    \nn \\& \hspace{1cm}
    Y_{\ell m}^*(\hat{\kv}_{1}) Y_{\ell m}(\hat{\xv}_{3}) Y_{J_1 j_1}(\hat{\xv}_{13}) Y_{J_1 j_1}^*(\hat{\xv}_{23})
    \nn \\&= \frac{1}{(2\pi)^6} \int d x_{13} x_{13}^2 \int d x_{23} x_{23}^2 \int d p_1 p_1^2 \int d p_2 p_2^2
    \nn \\& \hspace{0.5cm} \times 
    \sum_{ij} \sum_{\ell_6+\ell_7+\ell_8 \leq i} \sum_{\ell_9+\ell_{10}+\ell_{11} \leq j} \sum_{\ell_1, \ell_2, \ell_3, \ell_4, \ell_5, \ell_{12}, \ell_{13}, \ell_{14}} \sum_{\ell'_8, \ell'_{11}} \sum_{J_1, J_2, J_3} i^{\ell_1 + \ell_2 - \ell_{12} - \ell_{13}} j_{\ell_1}(p_1 x_{13}) j_{\ell_2}(p_2 x_{23})  
    \nn \\& \hspace{1cm} 
    j_{\ell_{12}}(k_1 x_{13}) j_{\ell_{13}}(k_2 x_{23}) \mathcal{F}^{(ij)}_{\ell_3 \ell_4 \ell_5 \ell_6 \ell_7 \ell_8 \ell_9 \ell_{10} \ell_{11}}(p_1, p_2) Q^{(i+j)}_{j_1 j_2 j_3}(x_{13}, x_{23}) x_{13}^i x_{23}^j 
    \nn \\& \hspace{1cm}
    \begin{pmatrix}
    J_2 & \ell_8 & \ell'_8 \\
    0 & 0 & 0
    \end{pmatrix}^2
    \begin{pmatrix}
    J_3 & \ell_{11} & \ell'_{11} \\
    0 & 0 & 0
    \end{pmatrix}^2
    \frac{(4\pi)^{15}}{N_{\ell_3 \ell_4 \ell_5} N_{\ell_6 \ell_7 \ell_9} N_{\ell_{10} J_1 0}} \sum_{m, m_i, m'_i,j_i} \sum_{LM} (-1)^{m_3 + m'_8 + m'_{11} + j_1 + M} 
    \nn \\& \hspace{1cm}
    \mathcal{G}_{\ell_4 \ell L}^{m_4 m M}
    \mathcal{G}_{\ell_1 \ell_3 \ell_4 \ell_6 \ell_9}^{m_1 m_3 m_4 m_6 m_9} \mathcal{G}_{\ell_2 \ell_3 \ell_5 \ell_7 \ell_{10}}^{m_2,-m_3 m_5 m_7 m_{10}}
    \mathcal{G}_{\ell_1 \ell_{12} \ell_6 \ell_7 \ell'_8 J_1}^{m_1 m_{12} m_6 m_7,-m'_8 j_1}
    \mathcal{G}_{\ell_2 \ell_{13} \ell_9 \ell_{10} \ell'_{11} J_1}^{m_2 m_{13} m_9 m_{10}, -m'_{11}, -j_1} 
    \nn \\& \hspace{1cm}
    \mathcal{G}_{L \ell_5 \ell'_8 \ell'_{11}}^{-M m_5 m'_8 m'_{11}} \mathcal{G}_{\ell \ell_{12} \ell_{14}}^{m m_{12}  m_{14}}Y_{\ell_{14} m_{14}}(\hat{\kv}_1) Y_{\ell_{13} m_{13}}^*(\hat{\kv}_2),
\end{align}
where we have used Eq.~\eqref{contraction_Y} in the last step.
Eq.~\eqref{wa-effects} is a special case of Eq.~\eqref{final-step-derivation} in the case of uniform window functions $W(\xv)=1$.

\section{Evaluation of the Bessel integrals}
\label{sec:eval_bessel_integrations}
We are interested in solving integrals of the following kind
\begin{align}
\label{the_bessel_integrals}
\int d x_{13} x_{13}^2 \int d x_{23} x_{23}^2 \int d p_1 p_1^2 \int d p_2 p_2^2 ~ j_{\ell_1}(k_1 x_{13})  j_{\ell_2}(k_2 x_{23}) j_{\ell_{3}}(p_1 x_{13}) j_{\ell_{4}}(p_2 x_{23}) \mathcal{F}(p_1, p_2) x_{13}^i x_{23}^j.
\end{align}
In the case of $\ell_3 = \ell_1 \pm n_1$ (and similary $\ell_4 = \ell_2 \pm n_2$) for some integers $n_1, n_2$ these integrals can be solved analytically. 
Consider the following building block of Eq.~\eqref{the_bessel_integrals}
\begin{align}
\label{matrix-A}
\mathcal{A}^n_{\ell, \ell'}(p,k) \equiv \int dx x^{2 +n} ~ j_\ell (p x) j_{\ell'} (k x).
\end{align}
In the case of $n=0$, the couplings of the main Eq.~\eqref{wa-effects} only allows $\ell'=\ell$. In this case Eq.~\eqref{matrix-A} is simply the orthogonality of spherical Bessel identity
\begin{equation}
\mathcal{A}^0_{\ell, \ell}(p,k) = \frac{\pi}{2 k^2} \delta_D(k-p).
\end{equation}

When $n=1$, the couplings in the main equation Eq.~\eqref{wa-effects}  only allows $\ell' = \ell \pm 1$. In this case we have~\cite{Beutler:2020evf}
\begin{align}
\mathcal{A}_{\ell, \ell \pm 1}^1(p,k)  \nn &=  
-\left(\frac{\partial^2}{\partial p^2 } + \frac{2}{p} \frac{\partial}{\partial p} - \frac{\ell (\ell+1)}{p^2} \right) \int dx x ~j_\ell(px) j_{\ell \pm 1}(kx) 
\nn \\&= 
-\left(\frac{\partial^2}{\partial p^2 } + \frac{2}{p} \frac{\partial}{\partial p} - \frac{\ell (\ell+1)}{p^2} \right) 
\left\{ \begin{array}{cc} 
 \frac{\pi}{2} \Theta \left( k - p \right) p^\ell k^{-\ell-2}& : \ell' = \ell+1 \\
\frac{\pi}{2} \Theta \left( p - k \right) k^{\ell-1} p^{-\ell-1}&  : \ell' = \ell-1 
\end{array} \right.
\nn \\&=
\left\{ \begin{array}{cc}
\pi (\ell+1)p^{\ell-1}k^{-\ell-2} \delta_D(p-k) + \frac{\pi}{2} p^\ell k^{-\ell-2} \delta'_D(p-k)& : \ell' = \ell+1 \\
\pi \ell p^{-\ell-2}k^{\ell-1} \delta_D(p-k) - \frac{\pi}{2} p^{-\ell-1} k^{\ell-1} \delta'_D(p-k)& : \ell' = \ell-1
\end{array}\right.,
\end{align}
where $\delta'_D(r) \equiv \partial_r \delta_D(r)$ and the following identity has been used (see e.g.~\cite{Assassi:2017lea,DiDio:2018unb})
\begin{equation}
-\left(\frac{\partial^2}{\partial p^2 } + \frac{2}{p} \frac{\partial}{\partial p} - \frac{\ell (\ell+1)}{p^2} \right) j_\ell(px) = x^2 j_\ell(px).
\end{equation}

In the case of $n=2$, the couplings in the main Eq.~\eqref{wa-effects} allows for $\ell' = \ell, \ell \pm 2$. For $\ell'=\ell$, we have
\begin{align}
\mathcal{A}^{2}_{\ell, \ell}(p,k)
\nn &=
-\left(\frac{\partial^2}{\partial p^2} + \frac{2}{p} \frac{\partial}{\partial p} - \frac{\ell(\ell+1)}{p^2}\right) \int dx x^2 j_\ell(px) j_\ell(kx)
\nn \\&=
\frac{\pi}{2 p^4} \Big( (\ell^2 +\ell -2) \delta_D(p-k) + 2 p \delta'_D(p-k) - p^2 \delta''_D(p-k)\Big),
\end{align}
where $\delta''_D(r) \equiv \partial^2_r \delta_D(r)$. For $\ell' = \ell \pm 2$, we have 
\begin{align}
&\mathcal{A}^{2}_{\ell, \ell \pm 2}(p,k)
\nn \\&=
-\left(\frac{\partial^2}{\partial p^2} + \frac{2}{p} \frac{\partial}{\partial p} - \frac{\ell(\ell+1)}{p^2}\right) \int dx x^2 j_\ell(px) j_{\ell \pm 2}(kx)
\nn \\&=
-\left(\frac{\partial^2}{\partial p^2} + \frac{2}{p} \frac{\partial}{\partial p} - \frac{\ell(\ell+1)}{p^2}\right) \int dx x^2 j_\ell(px) \left((2\ell+1 \pm 2)\frac{j_{\ell \pm 1}(kx)}{kx} - j_\ell(kx)\right)
\nn \\&=
-\left(\frac{\partial^2}{\partial p^2} + \frac{2}{p} \frac{\partial}{\partial p} - \frac{\ell(\ell+1)}{p^2}\right)
\left\{ \begin{array}{cc}
(2\ell+3) \Big( \frac{\pi}{2}\Theta(k-p) k^{-\ell-3}p^\ell - \frac{\pi}{2p^2} \delta_D(p-k)\Big)&: \ell' = \ell+2 \\
(2\ell -1) \Big(\frac{\pi}{2} \Theta(p-k) p^{-1-\ell} k^{\ell-2} - \frac{\pi}{2 p^2} \delta_D(p-k)\Big)&: \ell' = \ell-2
\end{array} \right.
\nn \\&=
\left\{ \begin{array}{cc}
\left(-\frac{(-2+\ell+\ell^2)\pi}{2 p^4} + (3+5\ell+2\ell^2)\pi p^{\ell-1}k^{-\ell-3} \right) \delta_D(p-k)  \\
\hspace{0.3cm} +\left(-\frac{\pi}{p^3} + \frac{\pi}{2}(3+2\ell) p^\ell k^{-\ell-3} \right) \delta'_D(p-k) + \frac{\pi}{2 p^2} \delta''_D(p-k)
&: \ell' = \ell+2\\
\left(-\frac{(-2+\ell+\ell^2)\pi}{2 p^4} + \ell(-1+2\ell)\pi p^{-\ell-2} k^{\ell-2} \right) \delta_D(p-k)  \\
\hspace{0.3cm} + \frac{(-2 + (1-2\ell)p^{2-\ell} k^{\ell-2})\pi}{2 p^3}\delta'_D(p-k) + \frac{\pi}{2 p^2} \delta''_D(p-k)
&: \ell' = \ell-2
\end{array} \right..
\end{align}
Now defining 
\begin{align}
p_1^2 \mathcal{A}^i_{\ell_1, \ell_3}(p_1, k_1) &\equiv a_0(p_1,k_1) \delta_D(p_1 - k_1) + a_1(p_1,k_1) \delta'_D(p_1 - k_1) + a_2(p_1,k_1) \delta''_D(p_1 - k_1),
\nn \\ 
p_2^2 \mathcal{A}^j_{\ell_2, \ell_4}(p_2, k_2) &\equiv b_0(p_2,k_2) \delta_D(p_2 - k_2) + b_1(p_2,k_2) \delta'_D(p_2 - k_2) + b_2(p_2,k_2) \delta''_D(p_2 - k_2),
\end{align}
for some functions $a_0(p,k), a_1(p,k), a_2(p,k), b_0(p,k), b_1(p,k), b_2(p,k)$, the integral Eq.~\eqref{the_bessel_integrals} for the cases that we considered before, namely $(n=0, \ell'=\ell)$, $(n=1, \ell'=\ell \pm 1)$, $(n=2, \ell' = \ell, \ell \pm 2)$, can be evaluated as follows
\begin{align}
\int d x_{13} &x_{13}^2 \int d x_{23} x_{23}^2 \int d p_1 p_1^2 \int d p_2 p_2^2 ~ j_{\ell_1}(k_1 x_{13})  j_{\ell_2}(k_2 x_{23}) j_{\ell_{3}}(p_1 x_{13}) j_{\ell_{4}}(p_2 x_{23}) \mathcal{F}(p_1, p_2) x_{13}^i x_{23}^j
\nn \\&=
\Big [ a_0 b_0 \mathcal{F} - (a_0 b'_1 \mathcal{F} + a_0 b_1 \partial_{p_2} \mathcal{F}) + (2 a_0 b'_2 \partial_{p_2} \mathcal{F} + a_0 b_2 \partial^2_{p_2} \mathcal{F}) - (a'_1 b_0 \mathcal{F} + a_1 b_0 \partial_{p_1} \mathcal{F})
\nn \\& \hspace{0.4cm}+ 
[(a'_1 \partial_{p_2} \mathcal{F} + a_1 \partial_{p_1} \partial_{p_2} \mathcal{F}) b_1 + (a'_1 \mathcal{F} + a_1 \partial_{p_1} \mathcal{F}) b'_1] 
\nn \\& \hspace{0.4cm}- 
[(a'_1 \partial^2_{p_2}\mathcal{F} + a_1 \partial_{p_1} \partial^2_{p_2} \mathcal{F})b_2 -2(a'_1 \partial_{p_2} \mathcal{F}+ a_1 \partial_{p_1} \partial_{p_2}\mathcal{F})b'_2]
\nn \\& \hspace{0.4cm}+
(2a'_2 b_0 \partial_{p_1}\mathcal{F} + a_2 b_0 \partial^2_{p_1} \mathcal{F})
\nn \\& \hspace{0.4cm}-
[(2a'_2 \partial_{p_1}\mathcal{F} + a_2 \partial^2_{p_1}\mathcal{F})b'_1 + (2a'_2 \partial_{p_1}\partial_{p_2}\mathcal{F} + a_2 \partial^2_{p_1}\partial_{p_2}\mathcal{F})b_1]
\nn \\& \hspace{0.4cm}+
[(2a'_2 \partial_{p_1}\partial^2_{p_2} \mathcal{F} + a_2 \partial^2_{p_1} \partial^2_{p_2} \mathcal{F})b_2 + 2(2a'_2 \partial_{p_1} \partial_{p_2} \mathcal{F} + a_2 \partial^2_{p_1}\partial_{p_2}\mathcal{F})b'_2]
\Big]_{p_1 = k_1, p_2 =k_2},
\end{align}
where for example $a'_1(p,k) \equiv \partial_p a'_1(p,k)$ and similarly for the other $a(p,k)$'s and $b(p,k)$'s function.

\section{Convergence test}
\label{sec:convergence_test}
\begin{figure}[h!]
    \includegraphics[width=0.99\textwidth]{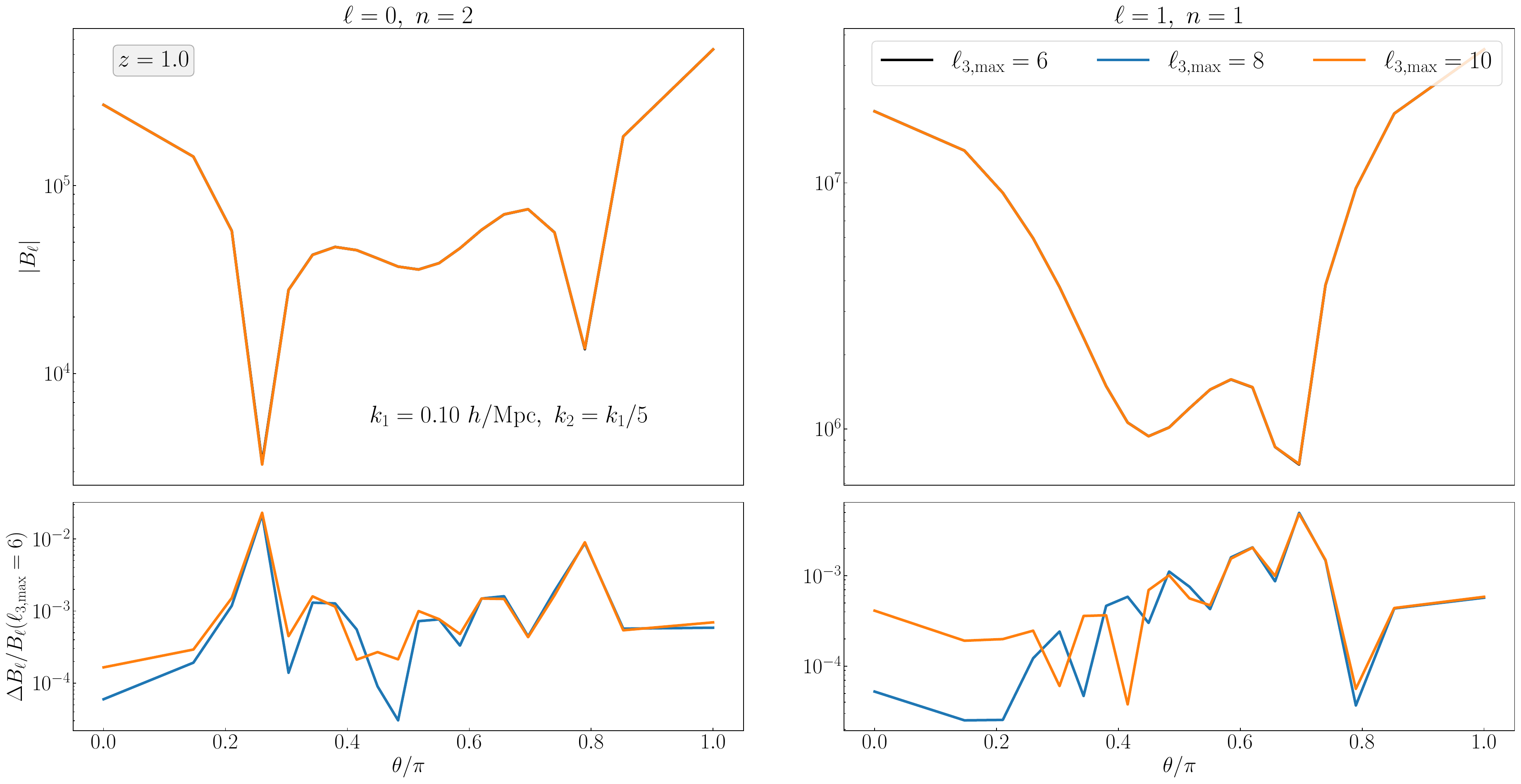}
    \centering
    \caption{Dependence of the results to the $\ell_{3,\mathrm{max}}$ which is the multipoles associated to $\hat{\pv}_1 \cdot \hat{\pv}_2$. The result is shown as a function of angle between the two wave-vectors $\theta$, where $\cos \theta = \hat{\mathbf{k}}_1\cdot \hat{\mathbf{k}}_2$. Increasing $\ell_{3,\mathrm{max}}$ to $\ell_{3,\mathrm{max}}=10$ gives a maximum percent level deviations with respect to the default $\ell_{3,\mathrm{max}}=6$ used in this work.}
    \label{fig:the_convergence}
\end{figure}

In this section we show the dependence of the results to the maximum value of $\ell_3$ (which is the multipoles associated to the angle between two wave-vectors of the bispectrum $\hat{\pv}_1 \cdot \hat{\pv}_2$). From Figure \ref{fig:the_convergence}, we found that a sufficiently large $\ell_{3,\mathrm{max}}=10$ gives a maximum percent level deviations for most triangle configurations with respect to the default $\ell_{3,\mathrm{max}}=6$ used in this work. Generally, the squeezed configurations required a smaller $\ell_{3,\mathrm{max}}$. 

\section{Flat-sky limit}

Here we compute the flat-sky limit of the bispectrum starting from Eq.~\eqref{wa-effects}. In flat-sky we have $i=j=0$ which implies $\ell_6= \ell_7 =\ell_8 = \ell_9 = \ell_{10} = \ell_{11}=0$. This leads to 
\begin{align}
\label{wa-effects-flatsky}
    B^{0}_\ell(k_1, k_2, &k_3) = 
    \nn \\& 
 \sum_{\ell_1, \ell_2, \ell_3, \ell_4, \ell_5, \ell_{12}, \ell_{13}} i^{\ell_1 + \ell_2 - \ell_{12} - \ell_{13}} 
    \nn \\& \times \Bigg[ \frac{1}{(2\pi)^6} \int \frac{d x_3 x_3^2}{V} \int d x_{13} x_{13}^2 \int d x_{23} x_{23}^2 \int d p_1 p_1^2 \int d p_2 p_2^2
    \nn \\& \hspace{1cm} 
    j_{\ell_1}(p_1 x_{13}) j_{\ell_2}(p_2 x_{23}) j_{\ell_{12}}(k_1 x_{13}) j_{\ell_{13}}(k_2 x_{23}) \mathcal{F}^{(ij)}_{\ell_3 \ell_4 \ell_5 000 000}(p_1, p_2)  \Bigg]
    \nn \\& \times \Bigg[
    \frac{(4\pi)^{14}}{N_{\ell_3 \ell_4 \ell_5} } \sum_{m, m_i} \sum_{LM} (-1)^{m_3  + M} \mathcal{G}_{\ell_4 \ell L}^{m_4 m M}
    \nn \\& \hspace{1cm}
    \mathcal{G}_{\ell_1 \ell_3 \ell_4 00}^{m_1 m_3 m_4 00} \mathcal{G}_{\ell_2 \ell_3 \ell_5 00}^{m_2 -m_3 m_5 00}
    \mathcal{G}_{\ell_1 \ell_{12} 000}^{m_1 m_{12} 000}
    \mathcal{G}_{\ell_2 \ell_{13} 000 }^{m_2 m_{13} 000} \mathcal{G}_{L \ell_5 00}^{-M m_5 00}
    \nn \\& \hspace{1cm}
    Y_{\ell_{12} m_{12}}^*(\hat{\kv}_1) Y_{\ell m}^*(\hat{\kv}_1) Y_{\ell_{13} m_{13}}^*(\hat{\kv}_2) \Bigg].
\end{align}
Then using the identities
\bea
\mathcal{G}^{\ell_1 \ell_2 \ell_3 0 0}_{m_1 m_2 m_3 0 0 } &=& \frac{\mathcal{G}^{\ell_1 \ell_2 \ell_3 }_{m_1 m_2 m_3  }}{4 \pi}
\nonumber \\
\mathcal{G}^{\ell_1 \ell_2 0 0 0}_{m_1 m_2 0 0 0 } &=& \frac{(-1)^{m_1}}{\left( 4 \pi \right)^{3/2}} \delta_{\ell_1 \ell_2} \delta_{m_1, -m_2}
\nonumber \\
\mathcal{G}^{\ell_1 \ell_2  0 0}_{m_1 m_2 0 0 } &=&\frac{(-1)^{m_1}}{4 \pi}  \delta_{\ell_1 \ell_2} \delta_{m_1, -m_2}
\eea
we have
\begin{align}
\label{exp_after_simplified_gaunts}
    B^{0}_\ell(k_1, k_2, &k_3) = 
 \sum_{\ell_1, \ell_2, \ell_3, \ell_4, \ell_5} 
    \Bigg[ \frac{1}{(2\pi)^6} \int \frac{d x_3 x_3^2}{V} \int d x_{13} x_{13}^2 \int d x_{23} x_{23}^2 \int d p_1 p_1^2 \int d p_2 p_2^2
    \nn \\& \hspace{1cm} 
    j_{\ell_1}(p_1 x_{13}) j_{\ell_2}(p_2 x_{23}) j_{\ell_{1}}(k_1 x_{13}) j_{\ell_{2}}(k_2 x_{23}) \mathcal{F}^{(ij)}_{\ell_3 \ell_4 \ell_5 000 000}(p_1, p_2)  \Bigg]
    \nn \\& \times \Bigg[
    \frac{(4\pi)^{8}}{N_{\ell_3 \ell_4 \ell_5} } \sum_{m, m_i}  (-1)^{m_3 + m_5} \mathcal{G}_{\ell_4 \ell \ell_5}^{m_4 m m_5}
    \mathcal{G}_{\ell_1 \ell_3 \ell_4 }^{m_1 m_3 m_4 } \mathcal{G}_{\ell_2 \ell_3 \ell_5 }^{m_2 -m_3 m_5 } \left( -1\right)^{m_1+m_2+m_5}
    \nn \\& \hspace{1cm}
    Y_{\ell_{1} -m_{1}}^*(\hat{\kv}_1) Y_{\ell m}^*(\hat{\kv}_1) Y_{\ell_{2} -m_{2}}^*(\hat{\kv}_2) \Bigg]
    \nn \\ =& 
 \sum_{\ell_1, \ell_2, \ell_3, \ell_4, \ell_5} 
    \Bigg[ \frac{1}{(4\pi)^4} \int \frac{d x_3 x_3^2}{V} 
    \mathcal{F}^{(ij)}_{\ell_3 \ell_4 \ell_5 000 000}(k_1, k_2)  \Bigg]
    \nn \\& \times \Bigg[
    \frac{(4\pi)^{8}}{N_{\ell_3 \ell_4 \ell_5} } \sum_{m, m_i}   \mathcal{G}_{\ell_4 \ell \ell_5}^{m_4 m m_5}
    \mathcal{G}_{\ell_1 \ell_3 \ell_4 }^{m_1 m_3 m_4 } \mathcal{G}_{\ell_2 \ell_3 \ell_5 }^{m_2 -m_3 m_5 } \left( -1\right)^{m_3}
    Y_{\ell_{1} m_{1}}(\hat{\kv}_1) Y_{\ell m}^*(\hat{\kv}_1) Y_{\ell_{2} m_{2}}(\hat{\kv}_2) \Bigg].
\end{align}
The sum over $\ell_1$ and $\ell_2$ run over the Gaunt factors and spherical harmonics only, therefore by using
\be
\sum_{\ell_3 m_3} \mathcal{G}_{\ell_1 \ell_2 \ell_3}^{m_1 m_2 m_3} Y_{\ell_3 m_3} = Y^*_{\ell_1 m_1} Y^*_{\ell_2 m_2}
\ee
we obtain
\begin{align}
    B^{0}_\ell(k_1, k_2, &k_3) = 
 \sum_{ \ell_3, \ell_4, \ell_5} 
 \int \frac{d x_3 x_3^2}{V} 
    \mathcal{F}^{(ij)}_{\ell_3 \ell_4 \ell_5 000 000}(k_1, k_2)  
    \nn \\& \times \Bigg[
    \frac{(4\pi)^{4}}{N_{\ell_3 \ell_4 \ell_5} } \sum_{m, m_i}   \mathcal{G}_{\ell_4 \ell \ell_5}^{m_4 m m_5}
 \left( -1\right)^{m_3}
    Y^*_{\ell_{3} m_{3}}(\hat{\kv}_1) Y^*_{\ell_{4} m_{4}}(\hat{\kv}_1)
    Y_{\ell m}^*(\hat{\kv}_1) 
    Y^*_{\ell_{3} -m_{3}}(\hat{\kv}_2)     Y^*_{\ell_{5} m_{5}}(\hat{\kv}_2) 
    \Bigg]
 \nn \\&
    = 
 \sum_{ \ell_3, \ell_4, \ell_5} 
 \int \frac{d x_3 x_3^2}{V} 
    \mathcal{F}^{(ij)}_{\ell_3 \ell_4 \ell_5 000 000}(k_1, k_2)  
    \nn \\& \times \Bigg[
    \frac{(4\pi)^{3}}{N_{0 \ell_4 \ell_5} } \sum_{m, m_i}   \mathcal{G}_{\ell_4 \ell \ell_5}^{m_4 m m_5}
    Y^*_{\ell_{4} m_{4}}(\hat{\kv}_1)
    Y_{\ell m}^*(\hat{\kv}_1) 
      Y^*_{\ell_{5} m_{5}}(\hat{\kv}_2) 
      \mathcal{L}_{\ell_3 }\left( \hat{\kv}_1 \cdot \hat{\kv}_2 \right)
    \Bigg]
     \nn \\&
    = 
 \sum_{  \ell_4, \ell_5} 
 \int \frac{d x_3 x_3^2}{V} 
    \mathcal{F}^{(ij)}_{ \ell_4 \ell_5 000 000}(k_1, k_2, \hat{\kv}_1 \cdot \hat{\kv}_2)  
     \nn \\& \times 
    \Bigg[
    \frac{(4\pi)^{3}}{N_{0 \ell_4 \ell_5} } \sum_{m, m_i}   \mathcal{G}_{\ell_4 \ell \ell_5}^{m_4 m m_5}
    Y^*_{\ell_{4} m_{4}}(\hat{\kv}_1)
    Y_{\ell m}^*(\hat{\kv}_1) 
      Y^*_{\ell_{5} m_{5}}(\hat{\kv}_2) 
    \Bigg] \, .
\end{align}
The sum over $m$ can be simplified as follows
\bea
&&
\sum_{m m_4 m_5}\mathcal{G}_{\ell_4 \ell \ell_5}^{m_4 m m_5}
    Y^*_{\ell_{4} m_{4}}(\hat{\kv}_1)
    Y_{\ell m}^*(\hat{\kv}_1) 
      Y^*_{\ell_{5} m_{5}}(\hat{\kv}_2) 
      \nn \\
      &=&\sum_{m m_4 m_5}
      \int d \hat {\mathbf{q}}
      Y_{\ell_{4} m_{4}}(\hat {\mathbf{q}})
    Y_{\ell m}(\hat {\mathbf{q}}) 
      Y_{\ell_{5} m_{5}}(\hat {\mathbf{q}}) 
    Y^*_{\ell_{4} m_{4}}(\hat{\kv}_1)
    Y_{\ell m}^*(\hat{\kv}_1) 
      Y^*_{\ell_{5} m_{5}}(\hat{\kv}_2) 
      \nn \\
      &=& \frac{N_{\ell \ell_4 \ell_5}}{\left(4 \pi \right)^3}
          \int d \hat {\mathbf{q}}
          \mathcal{L}_{\ell_4} \left( \hat {\mathbf{q}} \cdot \hat \kv_1 \right)                    \mathcal{L}_{\ell} \left( \hat {\mathbf{q}} \cdot \hat \kv_1 \right)
          \mathcal{L}_{\ell_5} \left( \hat {\mathbf{q}} \cdot \hat \kv_2 \right) 
               \nn \\
      &=& \frac{N_{\ell \ell_4 \ell_5}}{\left(4 \pi \right)^3} \sum_{\ell_1} \left( 2 \ell_1  +1 \right) 
 \begin{pmatrix}
\ell_4 & \ell & \ell_1\\
0 & 0 & 0
\end{pmatrix}^2
          \int d \hat {\mathbf{q}}
            \mathcal{L}_{\ell_1} \left( \hat {\mathbf{q}} \cdot \hat \kv_1 \right)
          \mathcal{L}_{\ell_5} \left( \hat {\mathbf{q}} \cdot \hat \kv_2 \right)      
                         \nn \\
      &=& \frac{N_{\ell \ell_4 \ell_5}}{\left(4 \pi \right)^2} 
 \begin{pmatrix}
\ell_4 & \ell & \ell_5\\
0 & 0 & 0
\end{pmatrix}^2
          \mathcal{L}_{\ell_5} \left( \hat \kv_1 \cdot \hat \kv_2 \right)    
\eea
which leads to
\begin{align}
    B^{0}_\ell(k_1, k_2, &k_3) 
    = 
 \sum_{  \ell_4, \ell_5} 
 \int \frac{d x_3 x_3^2}{V} 
    \mathcal{F}^{(ij)}_{ \ell_4 \ell_5 000 000}(k_1, k_2, \hat{\kv}_1 \cdot \hat{\kv}_2)  
     \nn \\& \times   
    4\pi\left( 2 \ell + 1\right)   
     \begin{pmatrix}
\ell_4 & \ell & \ell_5\\
0 & 0 & 0
\end{pmatrix}^2
          \mathcal{L}_{\ell_5} \left( \hat \kv_1 \cdot \hat \kv_2 \right)  
 \, .
\end{align}
In particular for the monopole $\ell=0$
\begin{align}
    B^{0}_0(k_1, k_2, &k_3) 
    =     4\pi \sum_{  \ell_4} 
 \int \frac{d x_3 x_3^2}{V} 
    \mathcal{F}^{(ij)}_{ \ell_4 \ell_4 000 000}(k_1, k_2, \hat{\kv}_1 \cdot \hat{\kv}_2)   
      \frac{ \mathcal{L}_{\ell_4} \left( \hat \kv_1 \cdot \hat \kv_2 \right)  }{2\ell_4 +1}
 \, .
\end{align}

\section{Useful identities}
\label{sec:useful_id}

In this appendix we collect several well-known mathematical identities employed in the derivation of the main results. Our convention for the spherical harmonics is such that 
\be
Y_{\ell 0}(\theta, \phi) = \sqrt{(2\ell+1)/4\pi} ~ \mathcal{L}_\ell(\theta)\,,
\ee 
where $\mathcal{L}_\ell(\theta)$ is Legendre polynomial of order $\ell$.

\paragraph{Rayleigh expansion of a plane wave:}
\begin{equation}
\label{plane_wave_exp}
    e^{i \vec{k} \cdot \vec{x}} = \sum_\ell i^\ell (2\ell +1) j_\ell(kx) \mathcal{L}_\ell (\hat{k} \cdot \hat{x})\,.
\end{equation}

\paragraph{Addition of spherical harmonics:}
\begin{equation}
\label{addition_spherical_harmonics}
    \mathcal{L}_\ell(\hat{x} \cdot \hat{y}) = \frac{4\pi}{2\ell +1} \sum_m Y_{\ell m} (\hat{x}) Y_{\ell m}^* (\hat{y})\,.
\end{equation}

\paragraph{Orthogonality of Legendre polynomials:} 
\begin{equation}\label{LL_ortho}
    (2\ell +1) \int \frac{d^2 \hat{k}}{4 \pi} \mathcal{L}_\ell(\hat{k} \cdot \hat{x}) \mathcal{L}_{\ell'}(\hat{k} \cdot \hat{y}) = \delta_{\ell \ell'} \mathcal{L}_\ell(\hat{x} \cdot \hat{y})\,,
\end{equation}
which implies that 
\begin{equation}\label{integrate_L_ell}
    \int \frac{d^2 \hat{k}}{4\pi} e^{i \vec{k} \cdot \vec{x}} \mathcal{L}_\ell(\hat{k} \cdot \hat{y}) = i^\ell j_\ell(kx) \mathcal{L}_\ell(\hat{x} \cdot \hat{y})\,.
\end{equation}

\paragraph{Gaunt's integral:}
\be
\label{integration_3Y}
    \int d^2 \hat{k}~ Y_{\ell_1 m_1} (\hat{k}) Y_{\ell_2 m_2} (\hat{k}) Y_{\ell_3 m_3} (\hat{k}) = \mathcal{G}^{m_1 m_2 m_3}_{\ell_1 \ell_2 \ell_3}\,.
\ee

\paragraph{Product of Legendre polynomials:} 
\begin{equation}\label{contraction_legendre}
    \mathcal{L}_{\ell_1} (\hat{k}\cdot \hat{x}) \mathcal{L}_{\ell_2} (\hat{k}\cdot \hat{x}) = \sum_{\ell_3} \begin{pmatrix}
\ell_1 & \ell_2 & \ell_3\\
0 & 0 & 0
\end{pmatrix}^2 (2\ell_3+1) \mathcal{L}_{\ell_3} (\hat{k}\cdot \hat{x}) \,.
\end{equation}

\paragraph{Product of spherical harmonics:}
\begin{equation}\label{contraction_Y}
    Y_{\ell_1 m_1} (\hat{n}) Y_{\ell_2 m_2} (\hat{n}) = \sum_{\ell_3, m_3} \mathcal{G}_{\ell_1 \ell_2 \ell_3}^{m_1 m_2 m_3} Y^*_{\ell_3 m_3} (\hat{n})\, .
\end{equation}

\setlength{\bibsep}{2pt plus 0.5ex}
\bibliographystyle{JHEP}
\bibliography{References}
\end{document}